\documentclass[a4paper,fleqn]{cas-dc}

\usepackage[numbers]{natbib}

\usepackage{microtype}

\def\tsc#1{\csdef{#1}{\textsc{\lowercase{#1}}\xspace}}
\tsc{WGM}
\tsc{QE}
\tsc{EP}
\tsc{PMS}
\tsc{BEC}
\tsc{DE}

\begin{document}
\let\WriteBookmarks\relax
\def\floatpagepagefraction{1}
\def\textpagefraction{.001}
\shorttitle{}

\shortauthors{Rabusov, Oeftiger, and Boine-Frankenheim}

\title[mode=title]{Characterization and minimization of the half-integer stop band with space charge in a hadron synchrotron}

\author[1]{Dmitrii Rabusov}
\ead{rabusov@temf.tu-darmstadt.de}
\address[1]{Technische Universit\"{a}t Darmstadt, Schlossgartenstr. 8, 64289 Darmstadt, Germany}
\cormark[1]

\author[2]{Adrian Oeftiger}%
\ead{a.oeftiger@gsi.de}
\address[2]{GSI Helmholtzzentrum  f\"{u}r Schwerionenforschung GmbH, Planckstr. 1, 64291 Darmstadt, Germany}

\author[1,2]{Oliver Boine-Frankenheim}
\ead{O.Boine-Frankenheim@gsi.de}

\begin{abstract}
In any hadron synchrotron, the half-integer resonance is among the strongest effects limiting the achievable maximum beam intensity. The heavy-ion superconducting synchrotron SIS100, currently under construction at GSI, should provide intense beams for the future FAIR experiments. Using SIS100 as an example, this paper develops a quantitative framework for characterizing the half-integer stop band for realistic, Gaussian-like distributed bunched beams. This study identifies the tune areas affected by the gradient-error-induced half-integer resonance for varying space charge strengths. A key insight of our analysis is that, for bunched beams a relatively small gradient error can result in a large half-integer stop band width. The achievable maximum bunch intensity, often referred to as space charge limit, is thus reduced significantly. This contrasts the findings in existing studies in literature based on more simplified beam distributions. The reason for discrepancy is identified in the increasing stop band width for Gaussian distributions when space charge becomes stronger, which appears on longer time scales as relevant for synchrotrons. The role of synchrotron motion in providing continuous emittance growth across the bunch is scrutinized. To minimize the half-integer stop band for a bunched beam, and hence increase the space charge limit, lattice corrections are applied: Including space charge in the optimization procedure recovers results equivalent to conventional lattice correction. Therefore, we find that conventional correction tools are well suited to increase the gradient-error-induced space charge limit of synchrotrons.

\end{abstract}


\maketitle

\section{\label{sec:introduction}Introduction}

Betatron resonances in a synchrotron can result in beam quality degradation, such as beam emittance growth and beam loss, and should be avoided during operation. With increasing resonance order, these undesired effects become weaker. The lowest order resonance affected by space charge is the half-integer (quadrupolar) resonance since dipolar resonances are independent of direct space charge. gradient errors are the source of the half-integer resonance.

First analytical studies on the half-integer resonance were carried out by Courant and Snyder~\cite{courant_et_snyder}. They performed the treatment based on single particle dynamics in the perturbed lattice by a gradient error. Though the stop band width is defined using this approach, it is done excluding collective effects. At the early stage of the design of high-intensity synchrotrons, space charge was realized to be a notable issue, addressing quadrupolar resonances. 

There is a large volume of published studies describing the role of space charge in coasting beams. Early examples of research into space charge include~\cite{Kapchinskij1959} and regarding the quadrupolar resonance with space charge~\cite{sacherer1968transverse}. More recent works like~\cite{machida1997} use the transverse Gaussian electric field of a coasting beam for stop band computations of the octupolar resonance. Also, several systematic investigations like~\cite{Warsop2016} and~\cite{cousenauLee} of a particle core model have been undertaken. The latter one has specifically targeted on the exact lattice structure and dispersion effects. The findings of~\cite{fedotov2002half} have confirmed in 2D particle-in-cell (PIC) simulations that the half-integer resonance occurs on the coherently shifted tune, and~\cite{Chao2015} elaborates on the envelope dynamics under space charge conditions and the stop band correction.

A large and growing body of literature has investigated the case of space charge in 3D distributions, since all synchrotrons operate with bunched beams. For example, bunched beams have been used in~\cite{machida1997} in simulations with frozen space charge model, and also in~\cite{fedotov2002half} in 3D PIC tracking with slow synchrotron motion but only for structure resonances. Recently, it has been shown in~\cite{Qiang2018} how to numerically compute stop bands of the envelope instabilities for bunched beams. The interplay between the incoherent and coherent effects in bunched beams has been described in~\cite{Hofmann2021}, though this work is focused on structure resonances. We shall indicate here that a comprehensive study of the quadrupole resonance with nonlinear space charge for realistic, i.e. 3D Gaussian-like distributed bunched beams, is lacking. The general case is challenging to tackle with analytical models, and this paper addresses this subject in self-consistent macro-particle tracking simulations.

\begin{figure}[t]
\includegraphics[width=\linewidth]{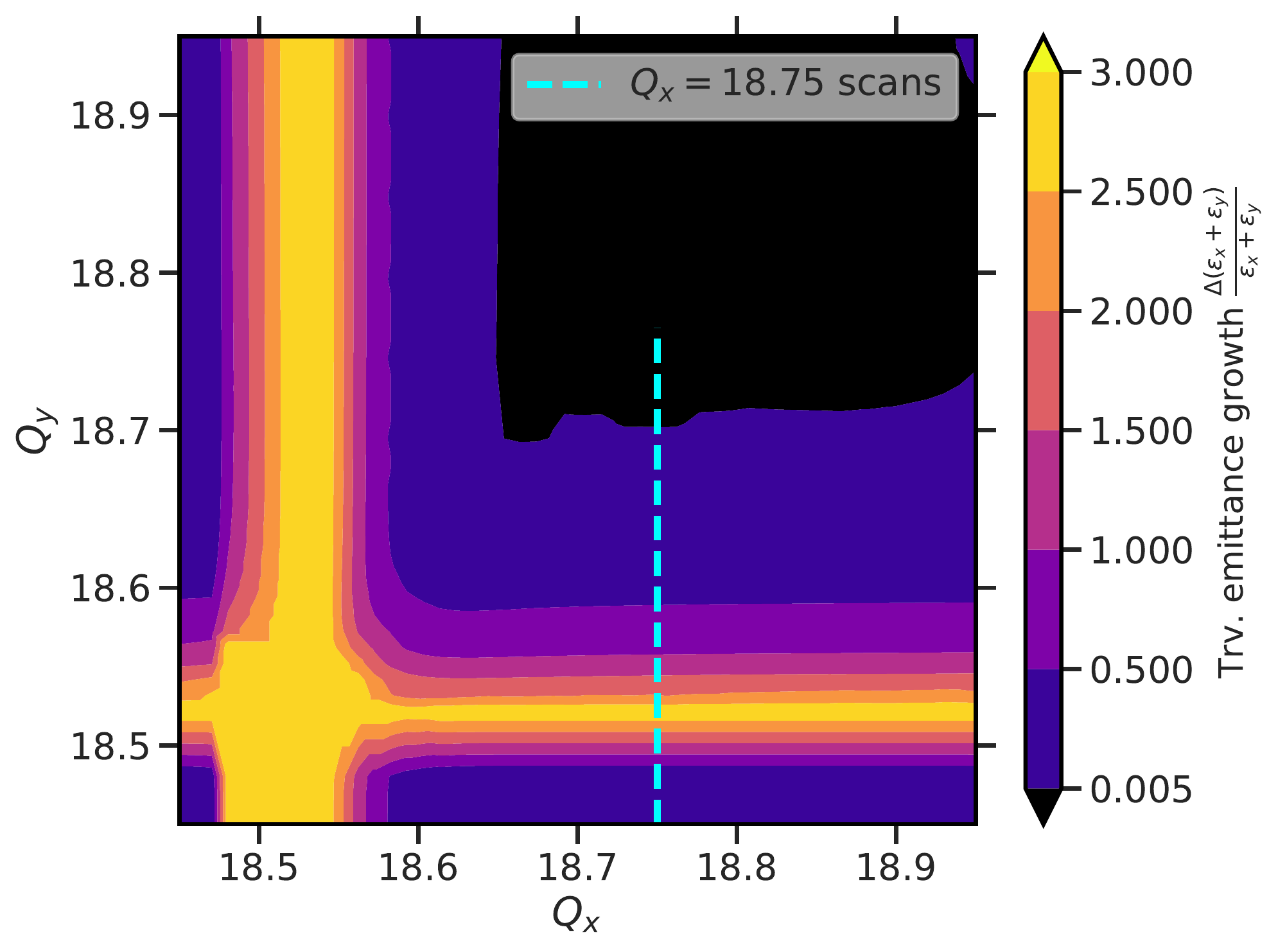}
\caption{\label{fig:tunescan} Transverse emittance growth over 200 turns vs. horizontal and vertical bare tunes in fully uncompensated SIS100 lattice.}
\end{figure}

As a relevant example case this study builds on the future SIS100 synchrotron featuring a gradient error. At the same time the synchrotron is to be operated at the space charge limit~\cite{Oeftiger2021}. Simulations of the uncompensated SIS100 scenario in Fig.~\ref{fig:tunescan} show significant emittance growth above the half-integer bare tunes due to the gradient error. These results demonstrate that the half-integer resonance is among the most important factors determining the available space in the tune diagram for high-intensity operation. The color scale indicates resonance-free zones (dark blue) and areas of the rapid emittance growth (yellow). Two thick yellow lines represent the stop bands associated with the horizontal and vertical quadrupolar resonances. They are located slightly above $Q_x,Q_y=18.5$ due to the defocusing effect from space charge.

The SIS100 model consists of a linear lattice with isolated gradient errors. The only nonlinear fields in the transverse plane are coming from the space charge potential of the Gaussian distributed beam. In this paper parameters to characterize the stop band width are space charge, the strength of the gradient error, and the synchrotron tune.

We organize this paper as follows. The approach of the half-integer stop band characterization in particle tracking simulations is shown in Sec.~\ref{sec:verification}. The main focus of the study is on simulations for bunched beams for varying space charge and gradient error strength. Coasting beam simulations are used to relate to previous results in literature. 
Next, Sec.~\ref{sec:longterm} confirms that the findings of the previous section hold for time scales as relevant for the synchrotron operation. Additionally, the influence of synchrotron tune on the total emittance growth inside the stop band is discussed. Then Sec.~\ref{sec:compensation} introduces the lattice correction to minimize the stop band width without space charge and proceeds to validate with 3D PIC space charge simulations that the correction also applies to bunched beams with space charge over long time scales. Finally, Sec.~\ref{sec:discussion} extrapolates the previous findings to predict the space charge limit in presence of gradient errors. Sec.~\ref{sec:conclusions} summarizes our major findings and concludes. 

\section{\label{sec:verification} Stop band characterization by tracking}

The following part of this paper describes in detail how the width of the half-integer resonance changes with varying space charge and a gradient error. Simulations are performed with coasting and bunched beams. The results for coasting beams are compared with theoretical treatises such as~\cite{sacherer1968transverse} and~\cite{Baartman}. Bunched beams are the subject of interest for synchrotrons, and are used here for realistic applications. Throughout this section, while comparing continuous and bunched beams we employ equal peak current conditions to obtain the same maximum strength of space charge in both cases. To start with, the simulation model and several parameters are introduced to quantify the strength of a gradient error and space charge.

\subsection{Gradient errors}

The beta-function $\beta(s)$ and the betatron phase advance $\mu(s)$ are functions of the path length $s$ and define the design optics of a synchrotron at the bare tune $Q$. The function $\Delta k(s)$ (gradient error) describes a deviation from the betatron focusing. Its strength can be quantified using stop band integrals,
\begin{equation}
    F_n = \frac{1}{2\pi} \oint \beta (s) \, \Delta k(s) \, \exp\left(-i \, n \, \frac{2\pi \, \mu(s)}{Q}\right) \, ds \,,
    \label{eq:stopband_integrals}
\end{equation}
first introduced by Courant and Snyder~\cite{courant_et_snyder}, corresponds to the half-integer stop band width at zero intensity. Gradient errors modify the transverse dynamics of the beam. A useful observable are the RMS beam envelopes, defined as the statistical beam moments $\sigma_x = \sqrt{\langle x^2 \rangle - \langle x \rangle^2} $ and $\sigma_y = \sqrt{\langle y^2 \rangle - \langle y \rangle^2} $ where the averaging $\langle \cdot \rangle$ is performed over a beam distribution. In the case of absent gradient errors and space charge, they have their design values expressed as
\begin{equation}
\begin{cases}
  \sigma_x(s) =  \sqrt{\epsilon_x \beta_x(s) + D(s)^2 \, \sigma_{(\Delta p / p)}^2} \\
  \sigma_y(s) = \sqrt{\epsilon_y \beta_y(s)} 
\end{cases}
\label{eq:beam_sizes}
\end{equation}
with the horizontal dispersion function D(s), where $\sigma_{(\Delta p / p)}$ is the RMS momentum spread. Usually, the vertical dispersion in synchrotrons like SIS100 is negligibly small. Therefore, we exclude it from our study. Also, we choose the vertical plane for probing the half-integer resonance at $Q_x = 18.75$ to isolate the gradient error resonance from other influences, for example, possible dispersion-related effects~\cite{OKAMOTO200265}. 

\begin{figure}[t]
\includegraphics[width=\linewidth]{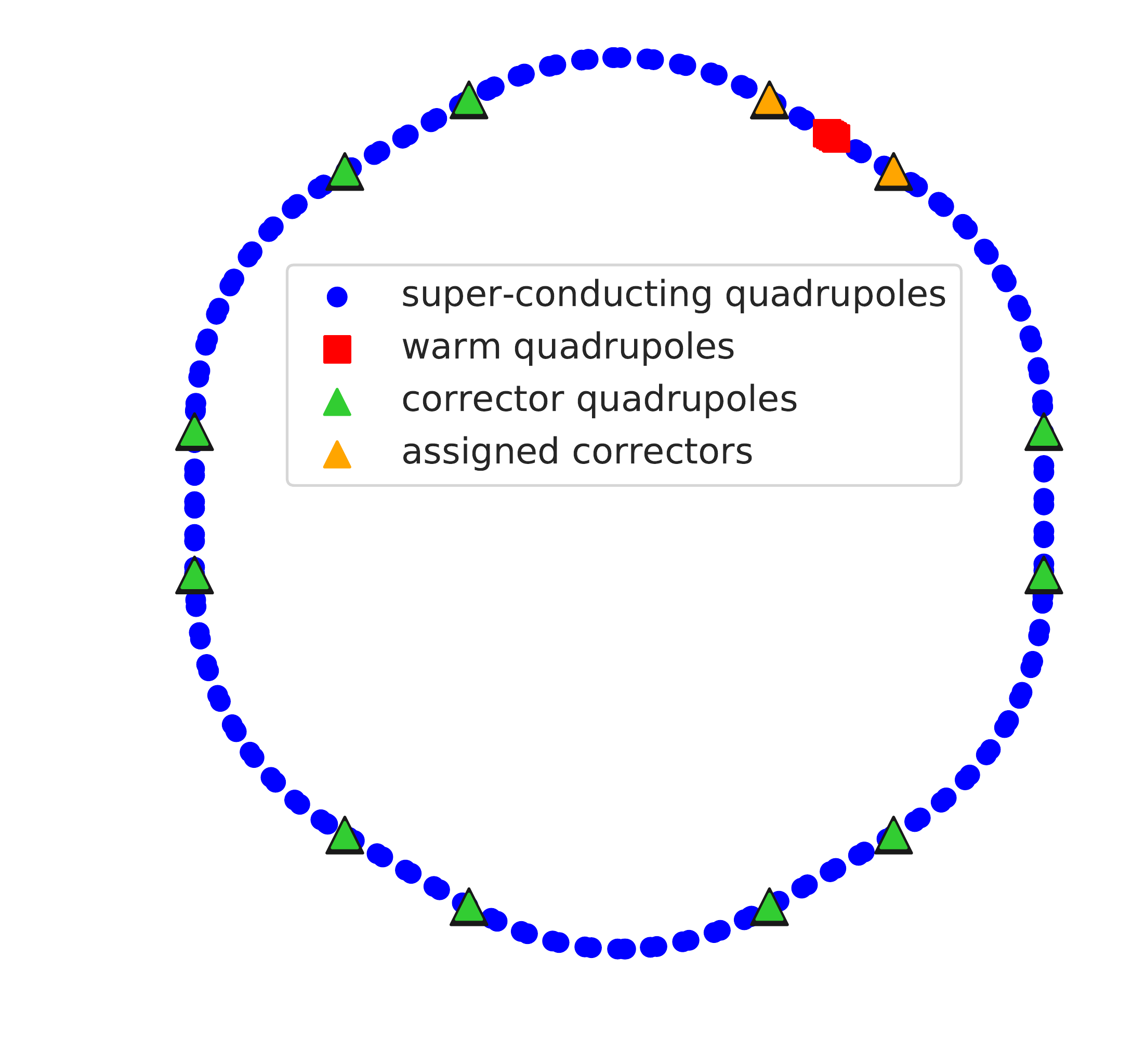}
\caption{\label{fig:survey}Survey of SIS100 gradient error setup.}
\end{figure}

A survey of the quadrupole magnets in SIS100 is shown in Fig.~\ref{fig:survey}. In our study case, the initial gradient error is given by a doublet of radiation-hard normal-conducting quadrupole magnets (red squares). These ``warm'' quadrupoles replace two superconducting ``cold'' magnets (blue circles) and thus supply weaker focusing strength. While the integral focusing strength is restored by an increased length of the warm magnets, this setup by design breaks the symmetry of the lattice. The detailed lattice model is described in~\cite{Oeftiger2021}. We control the gradient error using two quadrupole corrector magnets (orange triangles) located close to the warm quadrupole doublet. 

\begin{figure}[b]
\includegraphics[width=\linewidth]{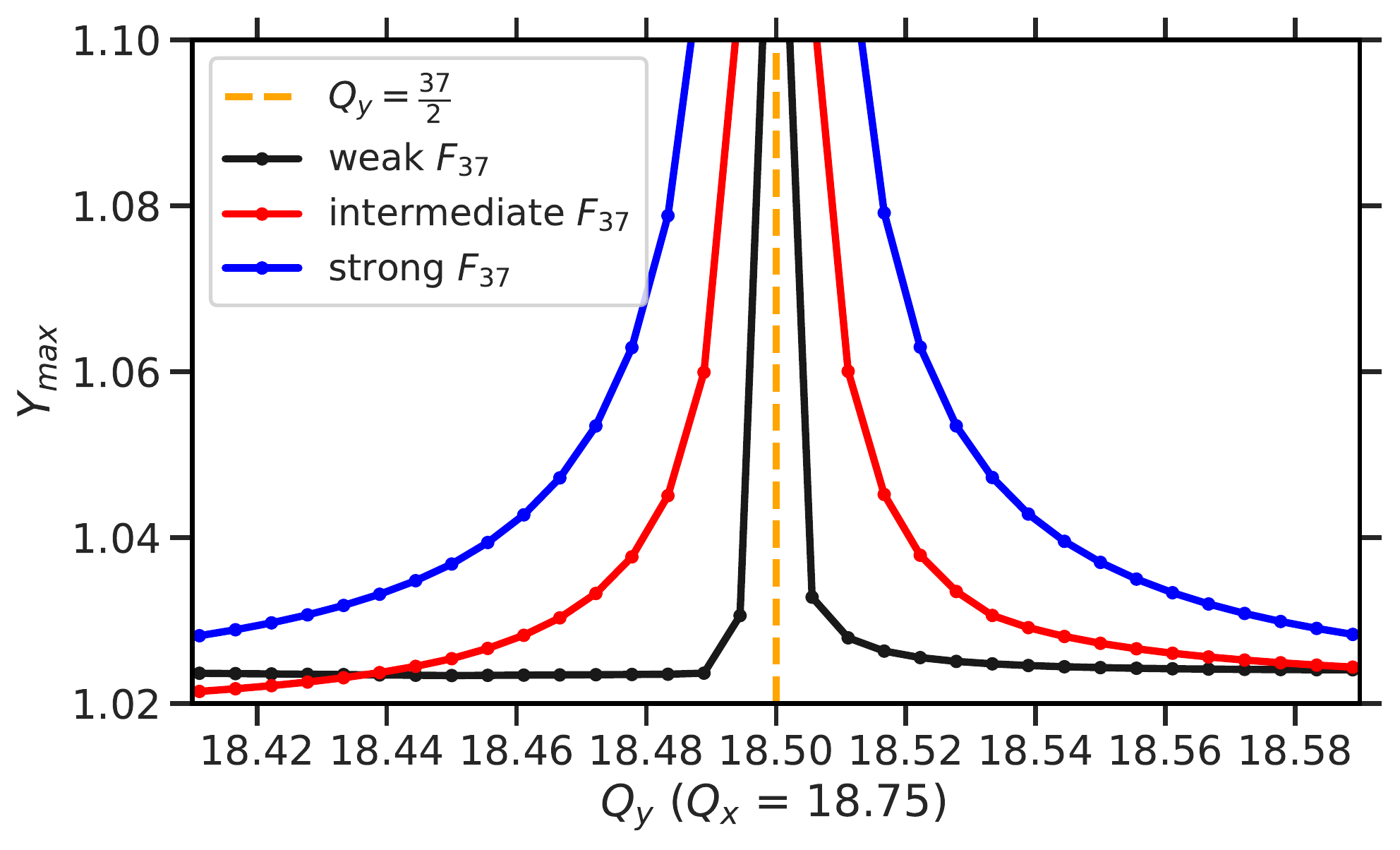}
\caption{\label{fig:gradient_errors}Response of vertical RMS envelopes of a coasting beam at zero intensity to bare vertical tune for different gradient error scenarios which are used along the paper.}
\end{figure}

Three basic scenarios of gradient error strength are considered in our study: (1.) \emph{weak} gradient error with correctors powered near their optimal values (the optimization is discussed in a greater detail in Sec.~\ref{sec:compensation}), (2.) \emph{intermediate} gradient error due to corrector current reduced to $75\%$, and (3) \emph{strong} gradient error with corrector current at $50\%$. Fig.~\ref{fig:gradient_errors} illustrates the impact of these three gradient error scenarios with plots of the response curves $Y_{\mathrm{max}} = \mathrm{Max}[\sigma_y / \sqrt{\epsilon_y \beta_y}]$ in terms of matched envelopes vs $Q_y$ the vertical bare tune. $\mathrm{Max}[\cdot]$ indicates the maximum along the path length $s$. The horizontal bare tune remains fixed at $Q_x = 18.75$.

\begin{table}[b]
    \caption{Reference values of stop-band integral $F_{37}$ for three gradient error scenarios}
    \label{tab:gradient_erros}
    \renewcommand{\arraystretch}{1.2}
    \centering
\begin{tabular}{ c|c|c|c } 
 \hline\hline
  & weak & intermediate & strong \\
 \hline
 $F_{37}$  & $0.6\cdot 10^{-3}$ & $2.6\cdot 10^{-3}$ & $4.3\cdot 10^{-3}$ \\
  \hline\hline
\end{tabular}
\end{table}

The orange dashed line shows the location of the zero-intensity half-integer resonance line. Response curves with black, red, and blue colors depict different values of the gradient error strength. As we see, the vertical RMS envelope increases around the resonance condition at the bare tune of $Q_y=18.5$, and the stop band becomes wider for stronger gradient errors. This is the usual behavior which is observed close to the quadrupolar resonance at negligibly small intensities.

\subsection{Space charge}

Fig.~\ref{fig:tunescan} shows how space charge shifts the vertical half-integer resonance from $Q_y = 18.5$ to higher values of bare tunes. This phenomenon is studied analytically in the literature for coasting beams, like in~\cite{sacherer1968transverse} and~\cite{Lapostolle}. The analytical concepts described below are used in this paper to quantify (maximum) space charge strength for both coasting and bunched beam conditions.

In the case of uniformly charged coasting beams (known as Kapchinskij-Vladimirsky or KV for short~\cite{Kapchinskij1959}), the quadrupolar resonance is described by the pair of equations on the horizontal and vertical envelopes of the beam. This approach is generalized in~\cite{sacherer1968transverse} on the wide range of transverse distributions with ellipsoidal symmetry. As mentioned above the horizontal bare tune remains fixed at $Q_x = 18.75$ for this paper. The quadrupolar stop band at lower $Q_y < Q_x = 18.75$ bare tunes is defined by the equation on the vertical envelope $\sigma_y$ because the motion is only loosely coupled between vertical and horizontal planes. In the smooth focusing approximation via constant $k(s)$ and $\beta(s)$ this equation becomes
\begin{equation}
\sigma_y'' + (Q_y^2 - 2\,Q_y\,\Delta Q^y_{\mathrm{KV}})\sigma_y - \frac{\epsilon_y^2}{\sigma_y^3} = 0
\label{eq:envelope_eq}
\end{equation}
where the value $\Delta Q_{\mathrm{KV}}$ is the KV tune shift derived in~\cite{Kapchinskij1959},
\begin{equation}
    \Delta Q^y_{\mathrm{KV}} = -\frac{K_{\mathrm{sc}} \, R^2 }{4 \, \langle \sigma_y \rangle \, (\langle \sigma_x \rangle + \langle \sigma_y \rangle) \, Q_y}
    \label{eq:sc_tuneshift_kv}
\end{equation}
with averaging $\langle \cdot \rangle$ over $s$. The effective ring radius is $R$. Parameters such as the beam peak current $I$, the ion charge number $Z$, the proton elementary charge $e$, the vacuum permittivity $\epsilon_0$, the rest mass of the ions $m_0$, the speed of light $c$, and relativistic factors $\gamma_r$, $\beta_r$ are combined into the space charge perveance,
\begin{equation}
K_\mathrm{sc} = \frac{Z\, e \, I}{2 \, \pi \,\epsilon_0 \, m_0 \, (\gamma_r \, \beta_r \, c)^3}\, .
\label{eq:perveance}
\end{equation}

The KV tune shift grows linearly with increasing peak intensity (peak current) and vanishes at high energies because of relativistic Lorenz factors $\beta_r$ and $\gamma_r$ in the denominator. Within a tune distance of $\simeq \frac{1}{4} \Delta Q^y_{\mathrm{KV}}$ to $Q_y=Q_x=18.75$~\cite{Baartman}, the 1D approximation of envelope equations breaks. This is sufficiently far away from the half-integer resonance, and we can neglect this effect in our study. 

For zero intensity, the solution of Eq.~\eqref{eq:envelope_eq} results in the mode with the frequency at the doubled bare tune. Finite space charge reduces the frequency by $\Delta Q_{\mathrm{env}}$~\cite{sacherer1968transverse}. This envelope tune shift amounts to
\begin{equation}
    \Delta Q_{\mathrm{env}} = 2 \,  C \cdot \Delta Q_{\mathrm{KV}}
    \label{eq:sc_tuneshift}
\end{equation}
where transverse beam geometry determines the constant $C$~\cite{Baartman}. In the SIS100 example the beam has $2b=a$, hence $C=2/3$ is the condition for the envelope space charge tune shift. The envelope mode is in resonance at
\begin{equation}
    2 Q - \Delta Q_{\mathrm{env}} = n
    \label{eq:resonance_condition}
\end{equation}
for integer $n$, which defines the quadrupolar linear resonance condition. For example, the resonance condition is met at $Q_y=18.5$ at zero intensity, whereas for nominal parameters, the space charge shifted condition is met at a bare tune of $Q_y=18.6$. Nominal parameters for SIS100 ${}^{238}\mathrm{U}^{28+}$ heavy-ion beam at injection energy are listed in Table~\ref{tab:parameters}\footnote{The values of $\Delta Q^y_{\mathrm{KV}}$ and $\Delta Q_{\mathrm{env}}$ correspond to transverse geometric RMS emittances $\epsilon_x = 8.75 \, \mathrm{mm} \, \mathrm{mrad}$, $\epsilon_y = 3.75 \, \mathrm{mm} \, \mathrm{mrad}$, and an RMS momentum spread $\sigma_{(\Delta p / p)} = 0.45\cdot 10^{-3}$ with an RMS bunch length of $\sigma_z=13.2\,\mathrm{m}$, and 0.625 $\cdot 10^{11}$ ions of ${}^{238}\mathrm{U}^{28+}$ per bunch.}. Often the maximum tune shift due to space charge in a Gaussian distribution, $\Delta Q_{\mathrm{Gauss}}$, is indicated as space charge strength: for fixed RMS emittances it amounts to twice the KV tune shift, $\Delta Q_{\mathrm{Gauss}} = 2 \cdot \Delta Q_{\mathrm{KV}}$.

\begin{table}[b]
    \caption{Reference parameters of the nominal uranium bunch distribution for SIS100 used in simulations (see Table I in Ref.~\cite{Oeftiger2021})}
    \label{tab:parameters}
    \renewcommand{\arraystretch}{1.2}
    \centering
\begin{tabular}{ c|c } 
 \hline\hline
 Parameter & Value \\
 \hline
 Particle & ${}^{238}\mathrm{U}^{28+}$ \\
 Ring circumference & 1083.6 m \\
 2D transverse distribution & Gaussian ($3.5 \sigma$ truncated) \\ 
 Peak current & 1.45 A \\
 Energy & 200 MeV/u \\
 $\Delta Q_{\mathrm{KV}}^{y} $ & 0.15 \\
  \hline\hline
\end{tabular}
\end{table}

To study the trends at varying space charge strength, the beam intensity is varied from zero to the nominal value, corresponding to the envelope tune shift between $0 \leq \Delta Q_{\mathrm{env}} \leq 0.2$. To maintain the same space charge strength for both coasting and bunched beams, the line charge density is scaled to obtain the same peak current. 

\subsection{Simulation model}

The linear resonance condition in Eq.~\eqref{eq:resonance_condition} is valid for coasting KV beams. It is generalized in~\cite{sacherer1968transverse} for a wide range of transverse beam distributions of coasting beams. The RMS 3D envelope equations derived in~\cite{sacherer-rms} can be applied to the case of bunched beams. There is a particular interest regarding the extend of the resonance from the linear resonance condition towards the stop band edges for realistic bunched 3D Gaussian beams. To accomplish this, macro particle simulations are used in this paper. Previously, such simulations have been utilized to study the half-integer stop band for KV beams in~\cite{cousenauLee} and waterbag beams in~\cite{HofmannBook} for coasting beam conditions.

The numerical simulations setup consists of macro particle tracking through the accelerator lattice. The beam is represented by macro particles. Each macro particle carries the charge of a set of particles. A set of macro-particles is generated by sampling a bivariate Gaussian distribution for the phase space of each transverse plane, matched to linear optics. Bunched beam is generated with a longitudinal bivariate Gaussian distribution, whereas coasting beam features zero momentum spread.

The library SixTrackLib~\cite{sixtracklib} is utilized for symplectic single-particle tracking through the lattice with gradient errors. The direct space charge interaction is modeled by lumped kicks which are applied in short steps along the lattice. Space charge is computed using the  slice-by-slice or 2.5D particle-in-cell (PIC) solver with PyHEADTAIL tracking code\footnote{Each PIC node solves the transverse Poisson equation for 64 longitudinal slices along the bunch with $128 \times 128$ grid. We use $10^6$ macro particles with $10^3$ equidistantly located in the accelerator lattice space charge nodes}~\cite{Oeftiger:2672381}. Statistical beam moments (see e.g. Ref.~\cite{Lee2011}) are used to analyze the results of particle tracking,
\begin{equation}
\begin{cases}
\sigma_y^2 = \langle y^2 \rangle - \langle y \rangle^2  \\
\sigma_{y'}^2 = \langle y'^2 \rangle - \langle y' \rangle^2 \\
\sigma_{yy'} = \langle (y - \langle y \rangle) \, (y' - \langle y' \rangle) \rangle \\
\epsilon = \sqrt{\sigma_y^2 \sigma_{y'}^2 - \sigma_{yy'}^2}
\end{cases}
\label{eq:emittance_computation}
\end{equation}
where $\langle \cdot \rangle$ is the averaging over a distribution, the displacement $y$ and the angle $y' = dy / ds$ are the coordinates of particles in the vertical phase space.

\subsection{How to determine the stop band width}
\label{sec:stopbandwidth}

\begin{figure}[htp]
\includegraphics[width=\linewidth]{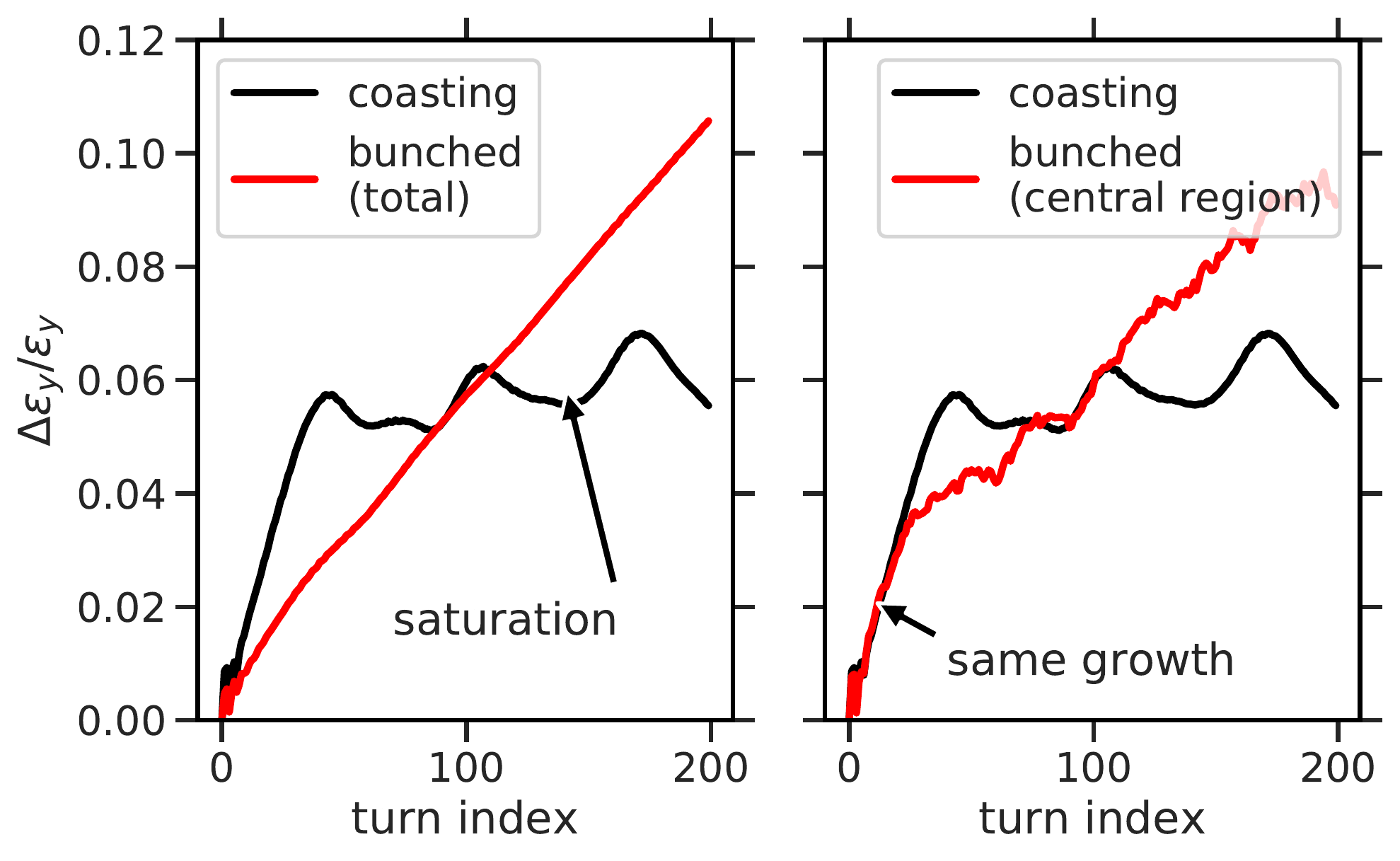}
\caption{\label{fig:on_resonance}The comparison of a coasting with a bunched beam in terms of the vertical emittance growth for the strong gradient error scenario. The working point is located slightly above the linear resonance condition.}
\end{figure}

First, we demonstrate how the resonant behavior can be observed in terms of the emittance growth. For illustrative purposes we use the case of the strong gradient error and space charge at nominal SIS100 beam parameters, i.e.\ the resonance condition is met at $Q_y  = 18.6$.  The bare tunes are set to  $Q_x = 18.75$ and $Q_y = 18.62$, slightly above the vertical quadrupolar linear resonance condition. Fig.~\ref{fig:on_resonance} displays typical responses to the half-integer resonance. The black curve is identical in both panels and indicates the vertical emittance growth for coasting beam conditions. 

The emittance fluctuates over the first $n \simeq 5$ turns before it proceeds to steadily increase linearly until $n\simeq 50$ turns. After that, the emittance reaches a saturation level. The red line on the left side of Fig.~\ref{fig:on_resonance} representing bunched beams shows us exactly the same behavior during the first $n \simeq 5$ turns. Afterward, the emittance increases at a lower rate than for the coasting beam. The red line on the right panel represents the central region of the bunched beam $|z| \ll \sigma_z$. Now the slope of the emittance growth is the same as for the coasting beam in $\simeq 30$ turns. We shall indicate here that the horizontal emittance under the same conditions stays practically constant (gaining less than $0.1 \%$ in 2000 turns). This confirms the horizontal and vertical planes are loosely coupled.

The dynamics during first five turns can be explained by the following. The initial transverse distribution is linearly matched to the local optics functions. The space charge potential distorts the phase space and this leads to the mismatch of the generated distribution, which turns to a rapidly decaying initial fluctuation in emittances. Next, the effect of emittance growth saturation appears in the case of coasting beams because of the decreasing space charge force as the beam size increases. This phenomenon has first been observed by Sacherer in~\cite{sacherer1968transverse} and is described in greater detail in~\cite{fedotov2002half}.

\begin{figure}
\includegraphics[width=\linewidth]{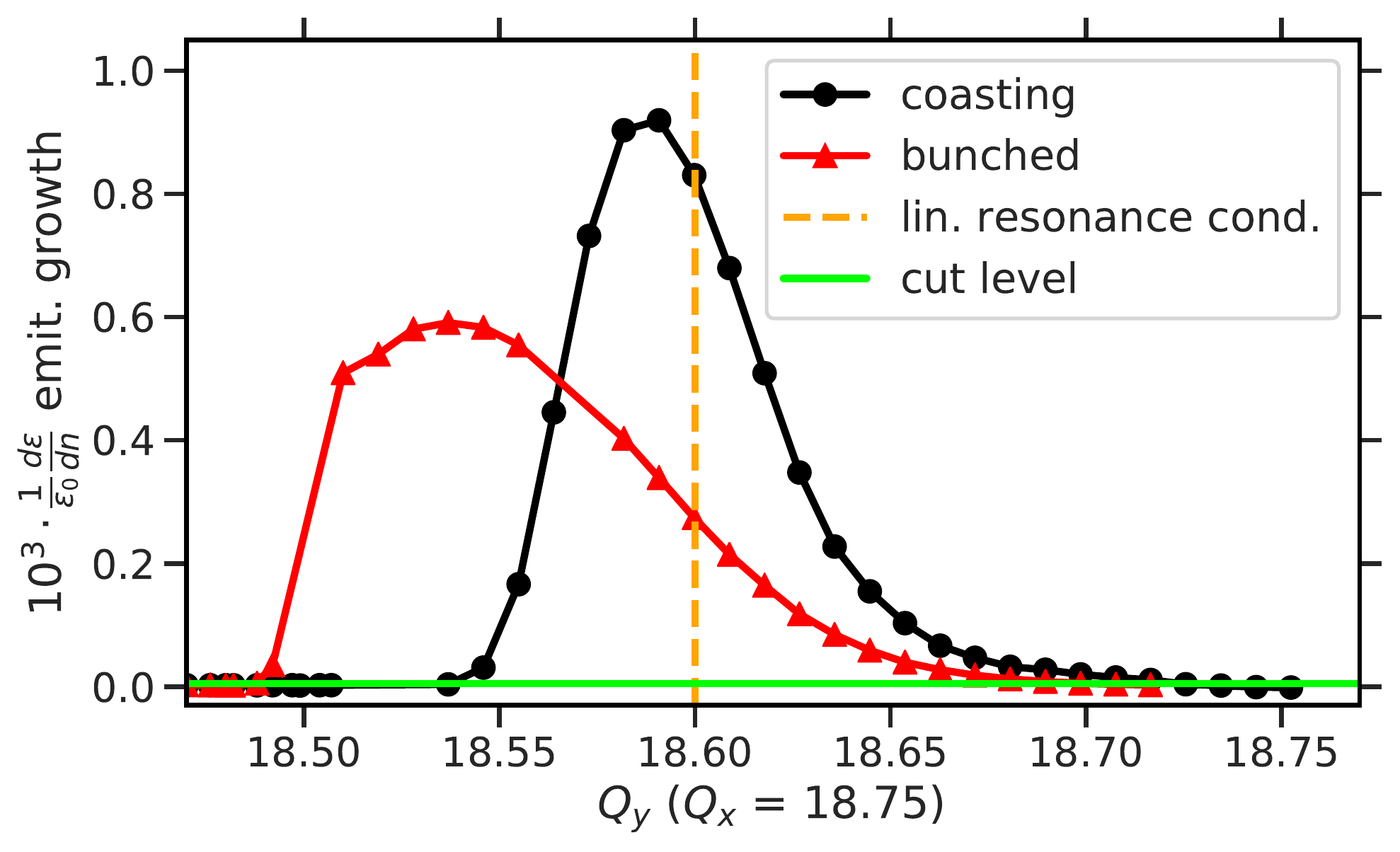}
\caption{\label{fig:coasting_vs_bunched}Example for the strong gradient error scenario how to get the stop band width using the threshold level from the vertical emittance growth rate.}
\end{figure}

Let us closely inspect in Fig.~\ref{fig:coasting_vs_bunched} the linear emittance growth regime between the initial mismatch and emittance saturation depending on the bare tune. 

With the black dots we represent the response curve for the coasting beam, whereas the red triangles show the corresponding curve for bunched beams. Both curves sharply rise and have their peaks between $18.47 < Q_y < 18.7$. Outside of this region the slope of emittance growth remains around a zero level marked by a solid lime green line. Although below the linear resonance condition $Q_y < 18.6$ depicted by the orange dashed line bunched beams have a wider range of tunes affected by the half-integer resonance compared to the coasting beams, above the orange dashed line coasting beams react to the same gradient error in a stronger way.
This difference is caused by varying longitudinal charge density in bunched beams. Let us consider the bare tunes above the orange dashed line. In the case of coasting beams, all longitudinal regions equally interact with the resonance, whereas in bunched beams only a fraction of particles (located in the bunch center) is involved in the emittance growth. That is why the shape of response curves significantly depends on the longitudinal beam distribution. 

The height of the solid lime green line determines the lower and upper edges of the stop band. Any working point with emittance growth above this threshold is considered as affected by the resonance. Here and throughout this work we use a threshold level of $\frac{1}{\epsilon_0}\frac{d\epsilon}{dn} = 5\cdot 10^{-6}$. This amounts to an overall emittance growth of $0.5\%$ during $10^3$ turns. For comparison, the injection plateau of SIS100 lasts around one second (or $1.6 \cdot 10^5$ turns), during which the beam emittance would grow by $80 \%$ according to this threshold. This may result in non-acceptable particle losses for typical apertures in synchrotrons. Hence, it is important first to identify the location and the width of the stop band, and then to compensate it. Another beneficial aspect of the developed technique of the stop band characterization is that it excludes the effect of the initial distribution mismatch. Though the mismatch can increase with space charge, it does not have any resonant nature. This means that any other technique which compares only final emittances has a systematic error in its design and always provides exaggerated results. Furthermore, the saturation effects are excluded when using the slope of the linear emittance growth. Hence, it is possible to adequately compare simulation results of bunched and coasting beams.

\subsection{Results: stop band width vs.\ space charge}

\begin{figure}[t]
\includegraphics[width=\linewidth]{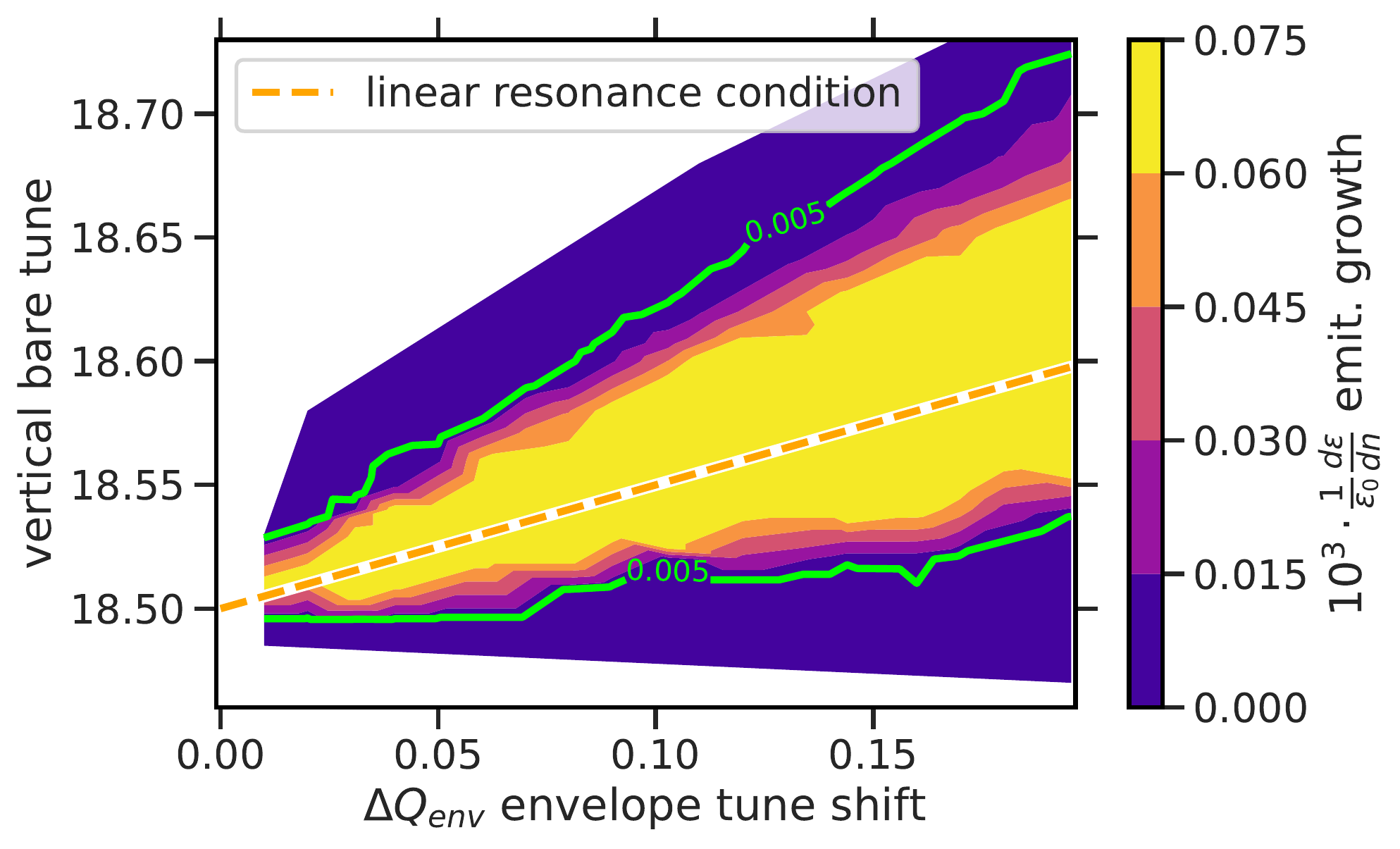}
\caption{\label{fig:coasting_anisotropic} Vertical emittance growth rate vs. space charge and the vertical bare tune for coasting beams, strong gradient error scenario is used.}
\end{figure}

The RMS envelope equations determine the quadrupolar resonance condition. This paper investigates the extent of the resonance towards the stop band edges. It is known (for example in~\cite{sacherer1968transverse} and~\cite{Lapostolle}, demonstrated in this paper in Appendix~\ref{appendix_KV}) that the stop band edges shift with space charge in parallel to the linear resonance condition in the case of coasting KV beams. The key results of this subsection show how, for a transversely Gaussian distributed coasting or bunched beam, the dependence of the stop band width on space charge significantly differs from the KV case. Figure~\ref{fig:coasting_anisotropic} displays the simulation results for 200 turns with Gaussian-like transversely distributed coasting beams, based on the strong gradient error scenario.

Yellow color represents the area with rapid emittance growth, whereas negligible emittance growth areas are shown in dark blue. Though the peak follows the linear resonance condition, the areas with the same emittance growth above and below the linear resonance condition are not parallel and become wider. In order to investigate how the stop band width changes with the gradient error strength and space charge, we use the technique of stop band width characterization described above. 

\begin{figure}[b]
\includegraphics[width=\linewidth]{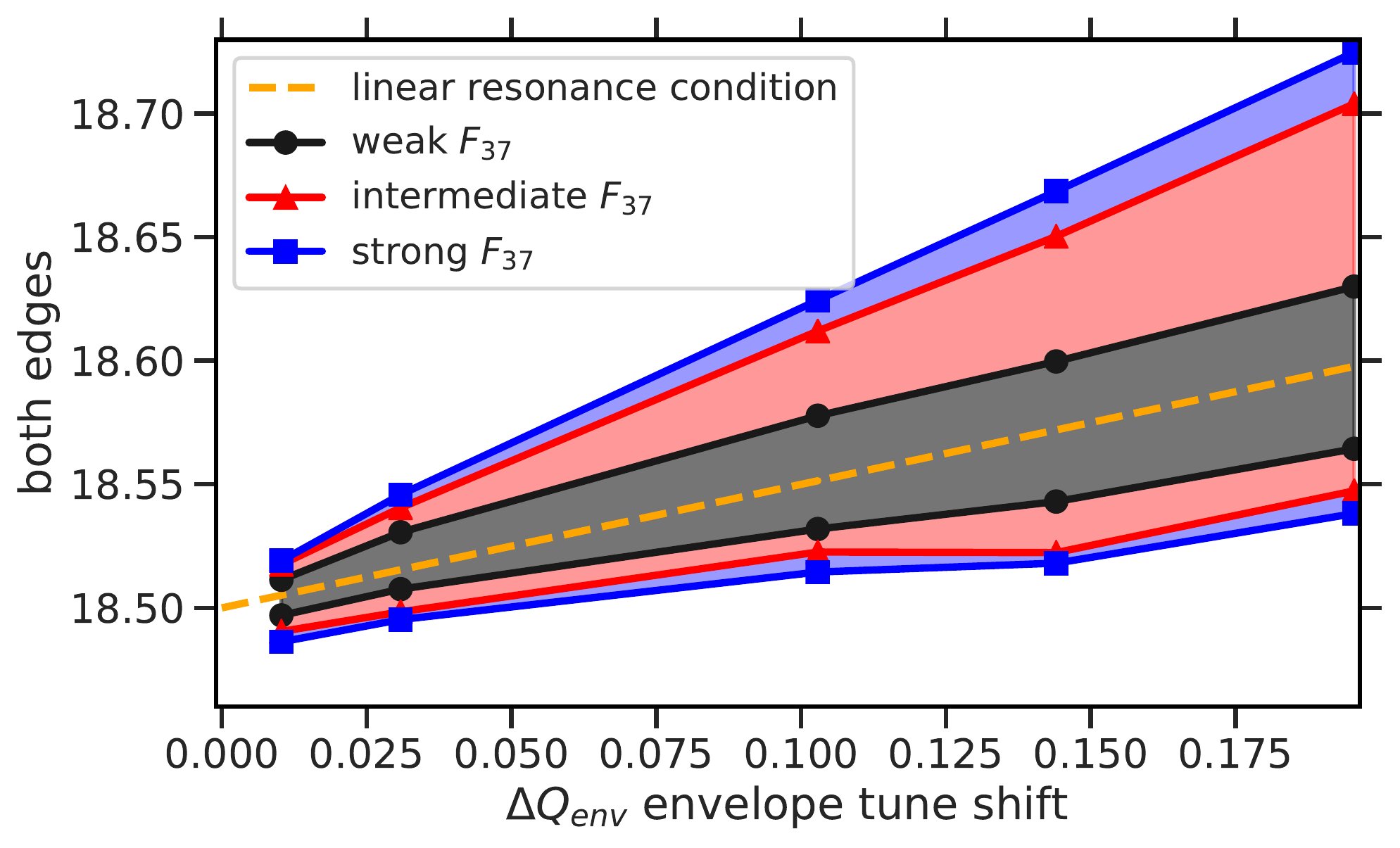}
\caption{\label{fig:characterization_2d} Stop band width for a coasting beam with varying gradient error and space charge.}
\end{figure}

Fig.~\ref{fig:characterization_2d} presents the coasting beams results. This plot shows the range of bare tunes affected by the resonance between lower and upper edges. As indicated by the blue squares, the strong gradient error leads to larger stop band values. Note, that both blue squares on the very right are the lower and upper edges in Fig.~\ref{fig:coasting_vs_bunched} for the coasting beam example. The area between the red triangles (corresponding to the intermediate gradient error case) is smaller than for the strong gradient error case. The smallest stop band width corresponds to the weak gradient error example depicted with the black dots. Fig.~\ref{fig:coasting_vs_bunched} quantitatively indicates how the half-integer resonance modifies with space charge. The edges of the stop band widen up with increasing space charge and are not parallel to the linear resonance condition for the whole range of intensities. This is observed for all probed gradient error strengths. The half-integer stop band width characterization technique fails when there is no space charge. The system becomes isolated, therefore transverse emittances are conserved. When determining the quadrupolar stop band for zero intensity, a single-particle approach can be used. In this scenario, the betatron motion of individual particles becomes unstable inside the stop band. The results of such analysis confirm the values listed in Table~\ref{tab:gradient_erros}.

\begin{figure}[b]
\includegraphics[width=\linewidth]{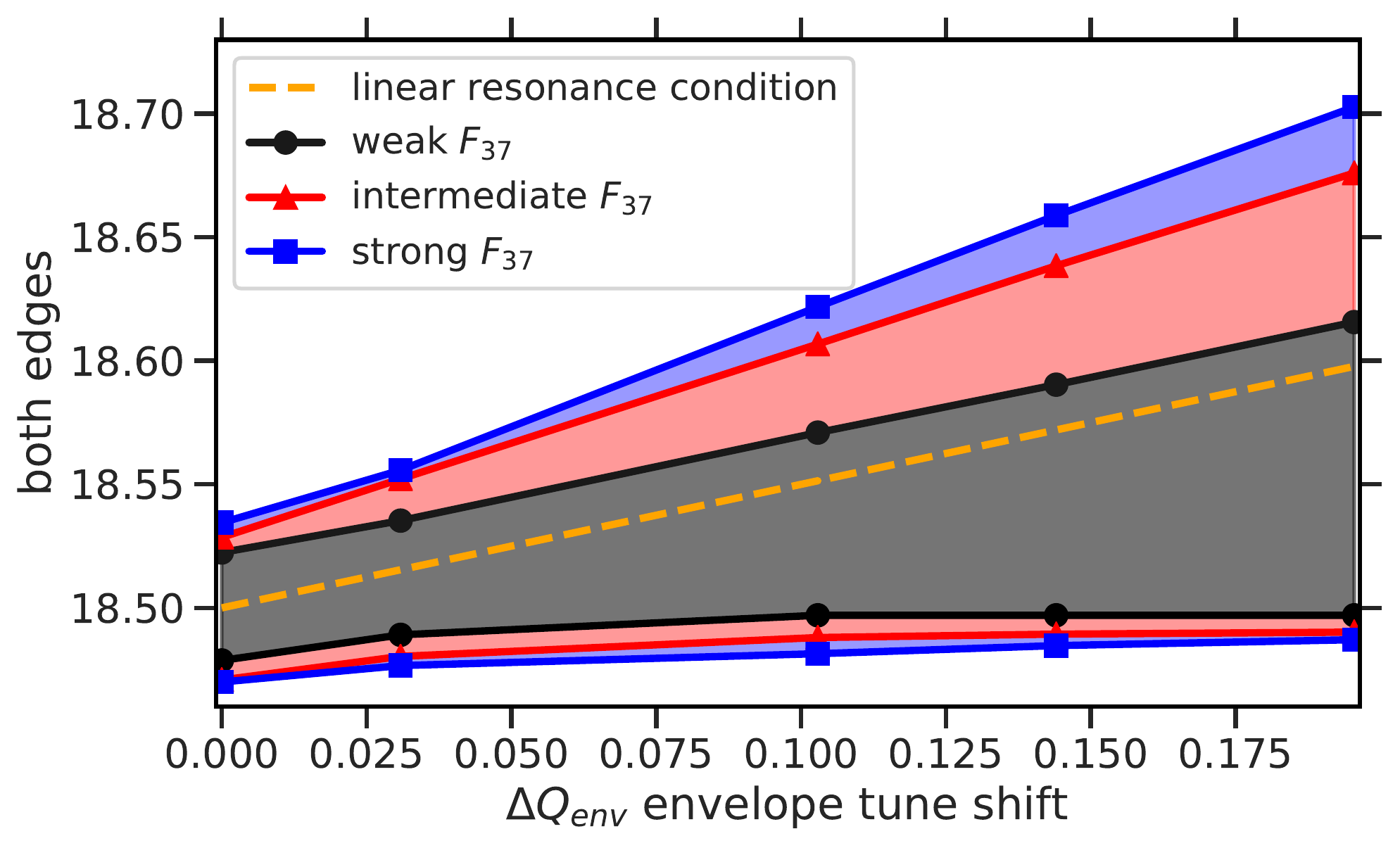}
\caption{Stop band width for a bunched beam with varying gradient error and space charge.\label{fig:edges}}
\end{figure}

Of more relevance for applications to synchrotrons is the situation for bunched beams. Employing the same stop band characterization approach, Fig.~\ref{fig:edges} depicts the simulation results. Again, like for coasting beams in Fig.~\ref{fig:characterization_2d}, the bunched beam stop bands (and the upper edge in particular) widen with increasing space charge for all three gradient error scenarios. Black dots which are the weak gradient error example constrain the smallest area. Red triangles and the blue squares represent the intermediate and strong gradient error cases, respectively. In comparison with coasting beams, the lower edge remains the almost same with increasing space charge. Any deviation in the lower edge is negligible compared with the upper edge change. Additionally, there is no significant correlation between the lower edge and the gradient error strength. Finally, the area occupied by the half-integer resonance expands due to the linear increase of the upper edge. The resonance free area for coasting beams between $18.5$ and the lower edge vanishes for bunched beams. This entails that, in the case of bunched beams, the stop band width can be reduced only by moving the upper edge down. In the weak space charge area where $\Delta Q_{\mathrm{env}} < 0.03$, the detuning from natural chromaticity plays a significant role. As a consequence, the stop band width at zero space charge is increased comparing to values in Table~\ref{tab:gradient_erros}. Space charge dominates in the area with $\Delta Q_{\mathrm{env}} > 0.03$, and the effect of the chromatic detuning becomes negligible.

Overall, this section has described the methods used in the characterization of the half-integer resonance. First, we have delineated the expressions to quantify the gradient error and space charge. Next, the setup of simulations and the stop band width characterization technique have been described. The technique is designed to adequately compare simulation results of coasting and bunched beams. Eventually, this technique has been applied to SIS100 yielding ranges of bare tunes affected by the half-integer resonance for various strengths of space charge and gradient error. We have presented how the stop band expands with increasing gradient error and space charge. And together these results provide important insights for such challenges as the lattice optimization and space charge limit in a synchrotron which follow in Sec.~\ref{sec:compensation} and in Sec.~\ref{sec:discussion} respectively.

\section{\label{sec:longterm}Long-term simulations}

The aim of this section is to demonstrate how the synchrotron motion affects the response to the half-integer resonance. Since for the majority of synchrotrons $Q_s$ lies between $0.01$ and $0.001$, and $Q_s / Q \ll 1$, we consider up to 2000 turns in this analysis. This corresponds to $\simeq 10$ synchrotron periods for the nominal SIS100 parameters.

A convenient way to change the synchrotron tune is to vary the RF voltage $V$ (see e.g.\ Ref.~\cite{Edwards1993}),
\begin{equation}
Q_s = \sqrt{- \frac{\eta \, h \, Z\,e\,V \, \cos\phi_s}{2\pi \, \beta_r^2 \, E}}
\label{eq:synchrotron_tune}
\end{equation}
where $\eta$ is the slip factor, $h$ is the harmonic number, $Z\,e$ is the ion charge, $\phi_s$ is the synchronous phase angle, $E$ is the beam energy. 

In order to keep the space charge conditions constant while varying the RF voltage $V$, bunch length and peak current are fixed. To match the varying RF bucket height, the longitudinal phase space distribution thus varies in rms momentum spread as
\begin{equation}
    \sigma_{\Delta p / p} = \sqrt{\frac{V}{V_{0}}} \cdot \sigma_{(\Delta p / p), 0}\, .
    \label{eq:momentum_scale}
\end{equation}

\begin{figure}[t]
\includegraphics[width=\linewidth]{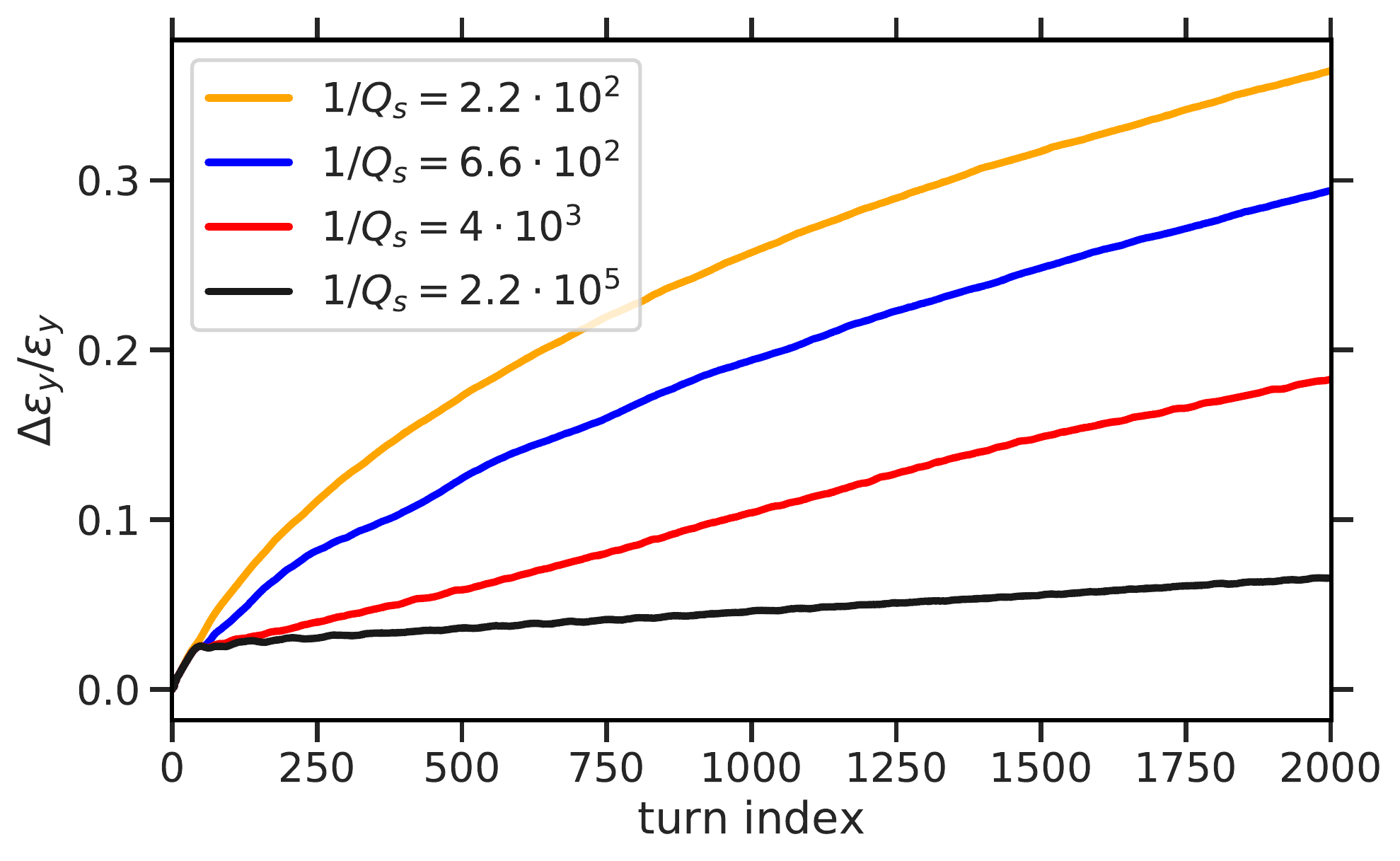}
\caption{Vertical emittance growth of Gaussian-like distributed bunched beam for various values of the synchrotron tune corresponding to the strong gradient error scenario.\label{fig:emittance_qs}}
\end{figure}

To remove the effect of chromatic detuning in our computer experiment, the beams are initialized with reduced transverse and longitudinal emittances.\footnote{To avoid bunch lengthening due to slippage effects and subsequent longitudinal drifting, the transverse emittances are scaled down along with the longitudinal emittance while maintaining the ratio $I / (\epsilon_x \cdot \epsilon_y)$ for the space charge tune shift to remain constant.} As a result, the space charge tune shift remains constant for varying $Q_s$ values. The example working point $Q_y = 18.62$ (same as in Fig.~\ref{fig:on_resonance}) serves to investigate the impact of $Q_s$. Given the strong gradient error scenario, there is significant emittance growth at this bare tune slightly above the linear resonance condition. Figure~\ref{fig:emittance_qs} illustrates the simulation results for various $Q_s$ values. Appendix~\ref{upper_long} repeats this analysis for the upper stop band edge where the driving term becomes very small.

The black line corresponds to relatively slow longitudinal motion with negligibly small $Q_s$. During the first 50 turns, the emittance increases linearly. After this, the speed of the linear growth significantly decreases. The red, blue, and orange lines correspond to increasing values of $Q_s$. Though the initial linear growth is the same for all of them, the total emittance growth for a given turn at later times, for instance $10^3$, increases with increasing synchrotron tune $Q_s$. It is interesting to note that, when relating to a given time in terms of synchrotron periods, the emittance growth per synchrotron period \emph{decreases} with increasing $Q_s$ (discussed in Appendix~\ref{upper_long}, Fig.~\ref{fig:qs_periods}).

The fastest growth rate corresponds to the initial short-term regime. After this, the trend is always slower than linear. The total emittance increase (via the quadrupolar resonance mechanism) during the injection plateau at the stop band edges is therefore less than the linearly extrapolated $80 \%$ (quoted in Sec.~\ref{sec:stopbandwidth}). Taken together, these results suggest that the developed scheme of the half-integer stop band characterization in Sec.~\ref{sec:verification} is valid also on long-term time scales: The tunes which are classified as resonance-affected remain inside the stop band, and the total resonantly gained emittance growth is always limited by the threshold.

\begin{figure}[b]
\includegraphics[width=\linewidth]{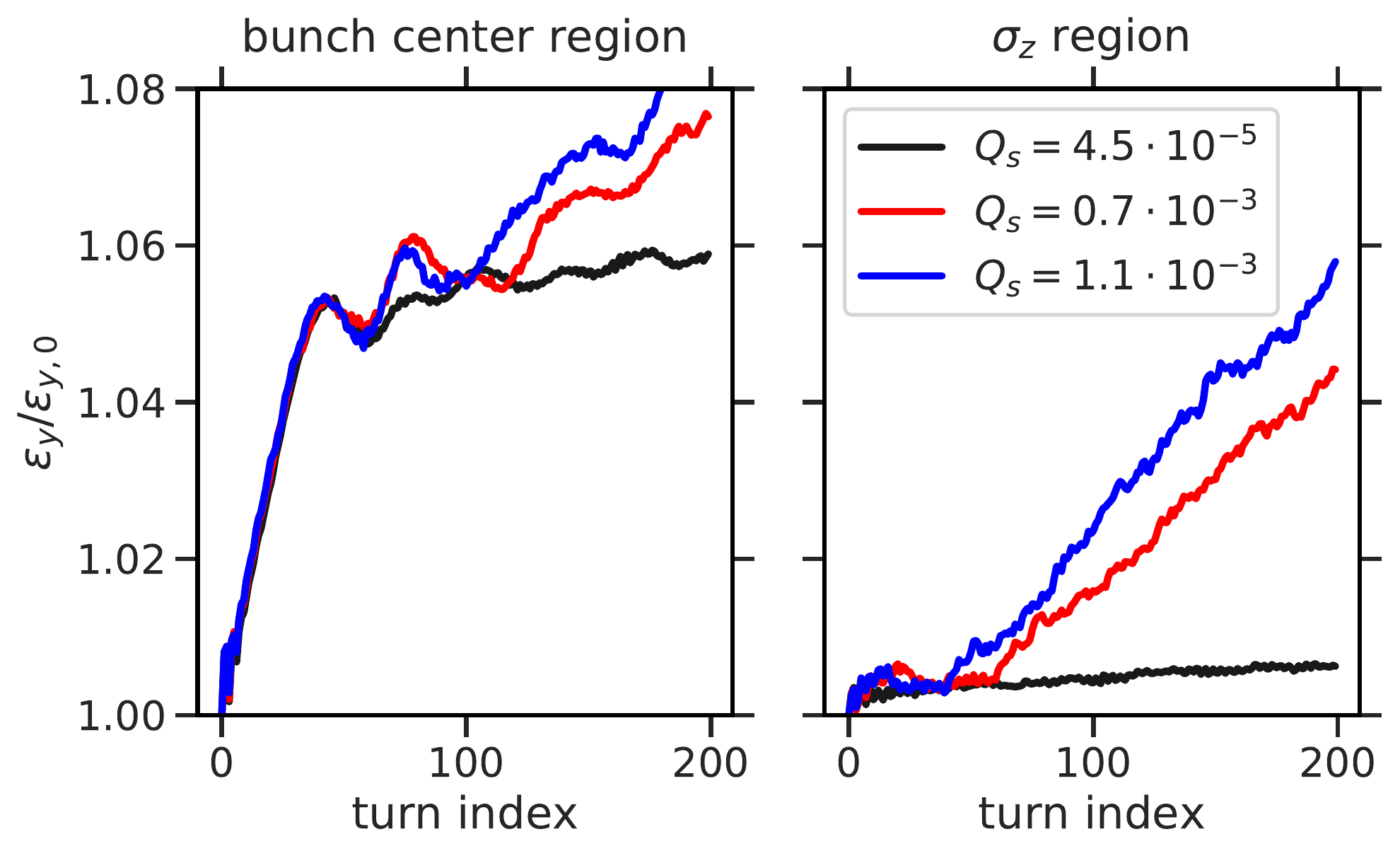}
\caption{Vertical emittance growth corresponding to separate longitudinal areas in the bunched beam for varying values of synchrotron tune\label{fig:qs_slices}. The working point is $Q_y = 18.62$ with the strong gradient error scenario}
\end{figure}

As indicated previously, in Fig.~\ref{fig:on_resonance} in Sec.~\ref{sec:stopbandwidth}, the emittance growth in bunched beams continues to increase while coasting beams saturate. Fig.~\ref{fig:qs_slices} displays how saturation is lost when $Q_s$ is greater than zero.

The curves show the simulated emittance growth at various synchrotron tunes, where the black color refers to $Q_s=4.5\times 10^{-5}$, red to $Q_s=0.7\times 10^{-3}$ and blue to $Q_s=1.1\times 10^{-3}$. The left panel corresponds to the bunch center region, whereas the right panel depicts the region towards the bunch ends, at a longitudinal position of $z=\sigma_z$. During the first 50 turns, the emittance in the $\sigma_z$ region remains relatively constant, whereas the bunch center resonantly reacts to the gradient error. After about 50 turns, a steady linear rise indicates emittance growth in the $\sigma_z$ region. 

The effect of synchrotron motion can be explained as follows: While particles with large amplitudes due to resonance interaction in the bunch center are transported towards the bunch ends, new particles from the bunch ends move towards the bunch center where they continue to interact with the resonance. This overall picture demonstrates the mechanism of interplay between the bunched beam and the half-integer resonance.

\section{\label{sec:compensation}Minimization of stop band}

So far, this paper has focused on the half-integer stop band width characterization on short and long time scales. The width of the half-integer stop band determines the flexibility of selecting the working point as well as the achievable maximum peak current (in terms of space charge tune shift) in a synchrotron. In this section, we discuss how to correct the lattice to achieve minimal stop band width. The applied lattice corrections are validated in tracking simulations with space charge.

\begin{figure}[b]
\includegraphics[width=\linewidth]{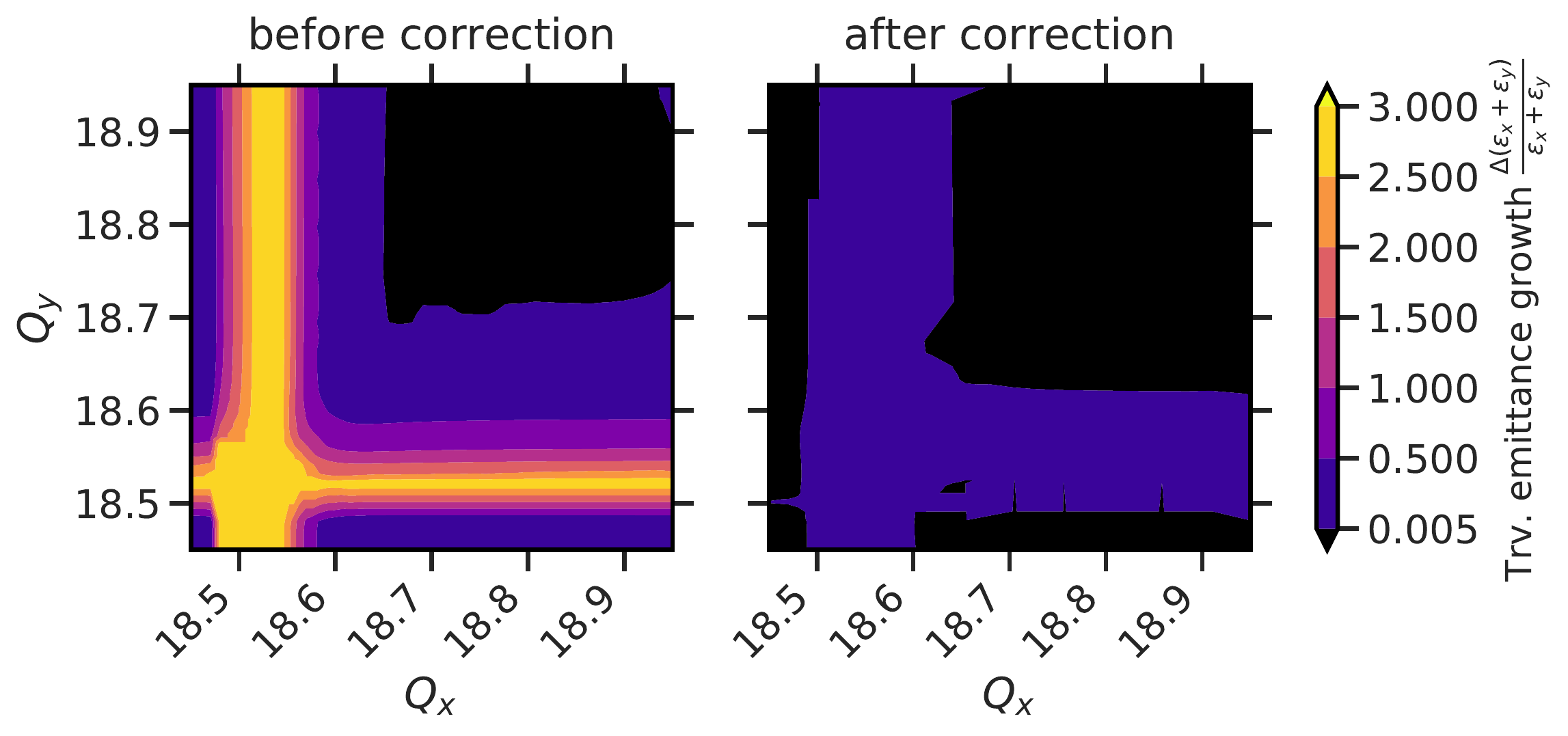}
\caption{\label{fig:before_after}Verification of the half-integer stop band minimization in terms of transverse emittance growth of a bunched beam.}
\end{figure}

In our SIS100 example, the initial gradient error is induced by the pair of warm quadrupoles. The integral focusing strength of warm quadrupoles can be independently adjusted. In addition, a pair of quadrupole corrector magnets located on either side of the perturbing warm quadrupoles is used to entirely suppress the beta-beating outside of the perturbation region~\cite{davidtwocorr}. Changing the integral focusing strength of the correctors and warm quadrupoles varies the stop band width. As shown previously, for instance in Fig.~\ref{fig:edges}, smaller values of $F_{37}$ lead to a narrower stop band also with space charge. Appendix~\ref{appendix_compensation} describes in detail the approach, exact integral focusing strength combinations, objective functions, and the numerical optimization scheme. The optimum configuration yields a stop band minimization in both transverse planes as demonstrated in Fig.~\ref{fig:before_after}.

The panels depict simulated emittance growth as a function of both transverse tunes for nominal bunched beam conditions during one synchrotron period. The left panel shows the simulation results without compensation (corresponding to Fig.~\ref{fig:tunescan} in Sec.~\ref{sec:introduction}) and the right panel with optimal compensation.

The color scale ranges from yellow for tunes affected by the resonance to dark blue for resonance-free tunes. This graph shows that the applied correction effectively suppresses the effects of the half-integer resonance in both directions for one synchrotron period.
This entails more freedom in choosing the working point or, alternatively, higher achievable bunch intensity. An important observation is that the optimal lattice configuration obtained without space charge is at the same time also the optimum configuration when including space charge for practical synchrotron applications. The case for SIS100 is demonstrated in Appendix~\ref{appendix_compensation} based on an envelope equation treatment.

\begin{figure}[t]
\includegraphics[width=\linewidth]{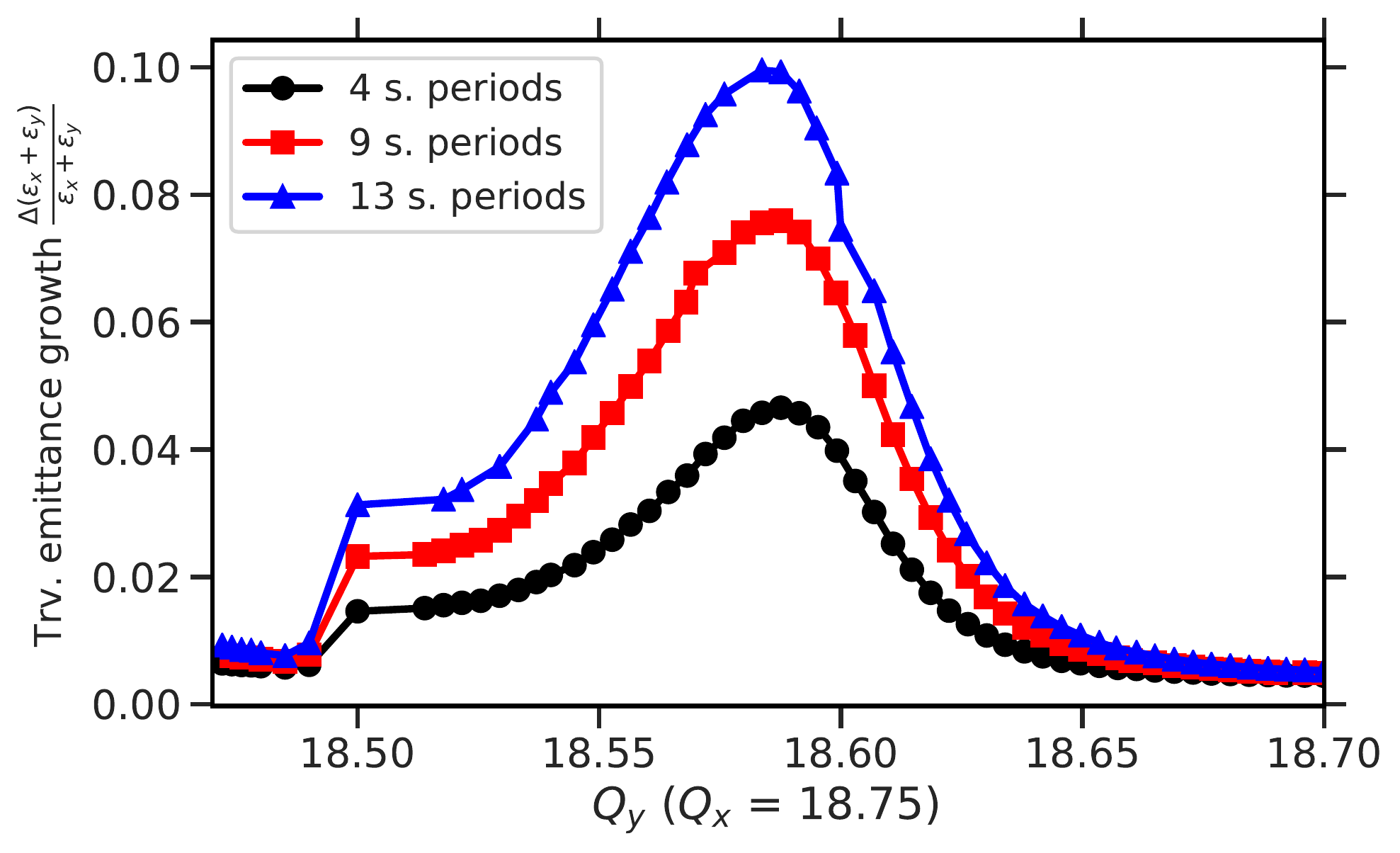}
\caption{\label{fig:response_3d_pic}Long-term transverse emittance growth for nominal SIS100 beam parameters after applying the lattice correction.}
\end{figure}

As shown in Sec.~\ref{sec:longterm} the emittance growth can cease only in the case of coasting beams or at negligibly small values of the synchrotron tune. Therefore, we perform a set of simulations at nominal beam parameters for more than 10 synchrotron periods to test the optimal settings on long-term time scales. Figure~\ref{fig:response_3d_pic} provides the summary for these simulations.

Starting with the black dots representing results after four synchrotron periods, we note that the half-integer resonance is still leading to the emittance growth. The red squares and blue triangles correspond to increasing numbers of turns. All the response curves sharply rise slightly below $Q_y = 18.5$, reach the peak around $Q_y = 18.6$, and then drop down. The highest value corresoponds to $10\%$ emittance growth which is reached after 13 synchrotron periods. This shows again (like in Sec.~\ref{sec:longterm}), that finite but small gradient errors with space charge lead to the finite emittance growth. Despite this, the overall inflicted emittance growth inside the stopband is suppressed compared to the non-compensated scenario. For example, without the correction, $\geq 10\%$ of the emittance growth is gained during one synchrotron period at any bare tune inside the $18.5 < Q_y < 18.6$ region, whereas now the beam reaches the same amount of the total emittance increase only around the linear resonance condition after 13 synchrotron periods.

This section has reviewed the three key aspects of the stop band minimization. First, the parameters for the lattice correction have been shown. Second, we have applied the optimal values in simulations and presented their performance. Finally, the validation in long-term simulations has been carried out. Though the half-integer resonance is not completely canceled, the overall emittance growth across the tune diagram is strongly reduced after the lattice correction.

\section{\label{sec:discussion}Space charge limit}

\begin{figure}[b]
\includegraphics[width=\linewidth]{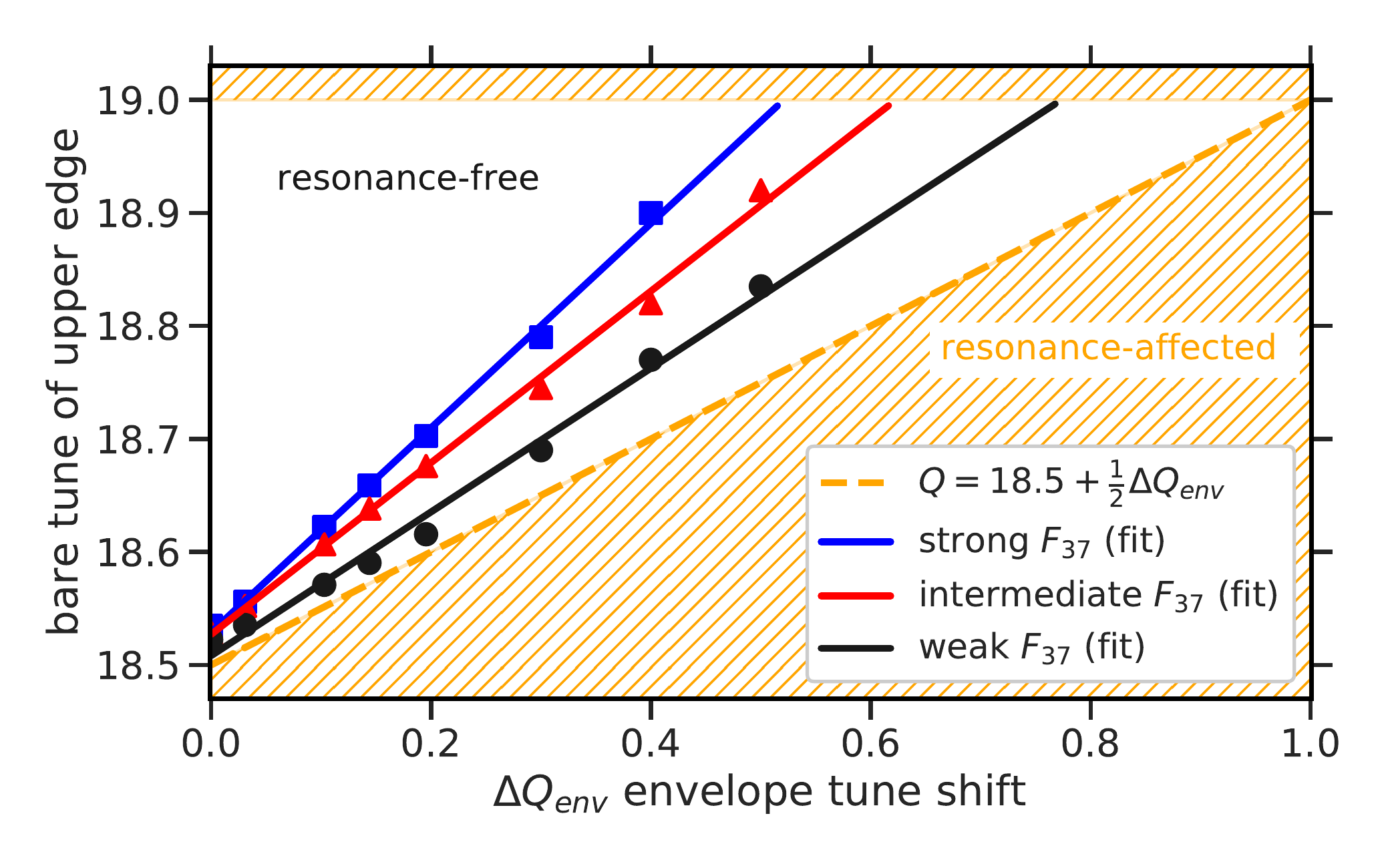}
\caption{\label{fig:space_charge_limit}Space charge limit. Blue squares, red triangles, and black dots are the locations of the upper edge corresponding to strong, intermediate, and weak gradient error scenarios from simulations}
\end{figure}

The aim of this section is to identify the space charge limit, i.e.\ the maximum tolerable space charge tune shift for realistic Gaussian-like distributed bunched beams where the area of available bare tunes (not affected by the quadrupolar resonance) shrinks to zero. Sec.~\ref{sec:verification} discusses the influence of space charge and gradient errors on the available area of bare tunes. These two separate influences are discussed here as degrees of freedom based on Fig.~\ref{fig:space_charge_limit}, which depicts the bare tune of the upper edge vs.\ the strength of space charge.

The classical conceptual discussion of the space charge limit only relates to the linear resonance condition, see e.g.\ Refs.~\cite{fedotov2002half, Baartman}: Given a finite but marginal gradient error, the space charge limit is reached when the linear resonance condition (shown in dashed orange) reaches the next half-integer (the horizontal line at $Q_y = 19$). Thus, the maximum achievable intensity is determined by the inequality
\begin{equation}
\Delta Q_{\mathrm{env}} < 1
\label{eq:coherent_limit}
\end{equation}
on the envelope tune shift. Next, the area below the orange dashed line becomes ``forbidden''. Therefore, the linear resonance condition sets the constraint on the bare tune $Q$ which shall satisfy the inequality $2Q > p + \Delta Q_{\mathrm{env}}$. 

The focus of this paper is on the extend of the quadrupolar resonance from the linear resonance condition towards the stop band edges for time scales relevant for a synchrotron. As it is described above, a finite gradient error moves the upper edge higher, leading to a steeper inclination of the curves in Fig.~\ref{fig:space_charge_limit} represented by blue squares (strong gradient error scenario), red triangles (intermediate gradient error), and black dots (weak gradient error). Note, that the points at $\Delta Q_{\mathrm{env}} < 0.2$ correspond to the upper edge curves in Fig~\ref{fig:edges}. In order to compute simulation results for $\Delta Q_{\mathrm{env}} > 0.2$, the effect of Montague resonance~\cite{montague1968fourth} is avoided by moving to higher horizontal bare tunes. Extrapolation to larger space charge strengths shows where the upper edge (colored solid lines) reaches the lower edge of the next higher $p+1$ half-integer resonance stop band, i.e.\ slightly below the horizontal line at $Q_y = 19$. This corresponds to the scenario where adjacent quadrupolar stop bands occupy the entire tune diagram. In our example case, the blue line meets the next resonance at lower space charge than the red line due to the stronger gradient error.

As it is shown above, a residual gradient error is not expected to result in relatively large emittance growth for coasting beams due to the saturation. Therefore, the space charge limit is reached at $\Delta Q_{\mathrm{env}} = 1$ (see e.g.\ Refs.~\cite{fedotov2002half, Baartman}). However, Sec.~\ref{sec:longterm} shows how the finite synchrotron motion always leads to significant emittance growth on the long term. Hence, the area below the blue (or red, black) lines in Fig.~\ref{fig:space_charge_limit} remains ``forbidden'' in the corresponding case. In other words, working points inside the stop band are not expected to conserve the emittances of bunched beams at finite synchrotron motion and a finite gradient error. This means that accurate estimations of intensity limitations in a synchrotron necessitate simulations with bunched beams. In application to the SIS100 example, the results obtained with Fig.~\ref{fig:space_charge_limit} are the following. According to the analytical expression in Eq.~\eqref{eq:coherent_limit}, the maximum possible intensity corresponds to $\Delta Q_{\mathrm{env}} = 1$, regardless of the gradient error strength. On the other hand, linear extrapolation of the simulation results including a certain gradient error indicates a maximum achievable intensity (in terms of $\Delta Q_{\mathrm{env}}$). The scenario of strong gradient error results in a limit of only $\Delta Q_{\mathrm{env}} \simeq 0.5$, the intermediate gradient error in $\Delta Q_{\mathrm{env}} \simeq 0.6$, and the weak gradient error in $\Delta Q_{\mathrm{env}} \simeq 0.8$. 

To conclude, we find that, for realistic Gaussian-distributed bunched beams, a relatively small stop band width at zero space charge ($\simeq 10^{-3}$, see Table~\ref{tab:gradient_erros}) can result in a significant reduction of the maximum intensity (here by a factor $2$ for the strong gradient error scenario). As a consequence, control and compensation of gradient errors (compare e.g.\ the strong to the weak scenario) is crucial for a synchrotron to maintain highest intensities under strong space charge conditions.

\section{\label{sec:conclusions}Conclusions and outlook}

Based on 2D and 3D simulation models with self-consistent space charge, this study characterizes the half-integer stop band for coasting and bunched beams in hadron synchrotrons. The FAIR heavy-ion synchrotron SIS100 serves as an example case, but our findings can be applied to other hadron synchrotrons. In our study, we compare the relevant stop band widths for varying gradient errors and space charge strengths. 

As one of the main findings, our stop-band characterization provides insights for choosing working points free of half-integer resonance impact. Coasting beam simulations establish the connection to existing analytical studies~\cite{sacherer1968transverse, Lapostolle}.
While the lower and upper edges of the stop-band for KV coasting beams are known to run parallel to the linear resonance condition, the stop-band width for Gaussian-distributed coasting beams is found to widen with increasing space charge. In this case, the upper edge of the half-integer stop band width linearly increases with space charge and gradient errors with a stronger slope than the linear resonance condition determined in~\cite{sacherer1968transverse}.

For bunched beams, simulations over several synchrotron periods in Sec.~\ref{sec:longterm} confirm that the stop-band characterization remains valid on longer time scales, relevant for realistic synchrotron operation. Periodic synchrotron motion is shown to result in continuous emittance growth. In SIS100, used here as an example case, choosing a working point outside the identified stop band ensures that the bunched beam is subject to an emittance growth below the chosen threshold during the injection plateau (here $80\%$ over one second).
Further, the influence of synchrotron tune $Q_s$ is investigated, with smaller $Q_s$ reducing the total emittance growth for a given time instant. As a key application we determine the intensity limit given by the half-integer stop band. We show that the correction of the half-integer resonance increases the intensity limit. The improvement is verified in long-term simulations.

Our simulation study for Gaussian bunches over long-term time scales (several synchrotron oscillations periods) resulted in several insights of practical relevance to existing and future synchrotrons. First, unlike in the case of coasting beams, the resonance-free area between the bare half-integer tune and the lower stop-band edge vanishes. Second, even a relatively small gradient error (resulting in a zero-space-charge stop-band width of just $\simeq 10^{-3}$) can considerably reduce the maximally achievable bunch intensity (in our example by up to a factor $\simeq 2$). We note that this effect---whilst absent in classical discussions of the space charge limit---must be taken into account under realistic synchrotron operation conditions. Third, the reduction of the half-integer stop band via lattice correction, computed without space charge, is found to also be optimal under finite space charge conditions. Therefore, conventional lattice correction tools are well suited to increase the gradient-error-induced space-charge limit of a synchrotron.

\section{Appendix}

\subsection{\label{appendix_KV}KV coasting beams}

\begin{figure}[t]
\includegraphics[width=\linewidth]{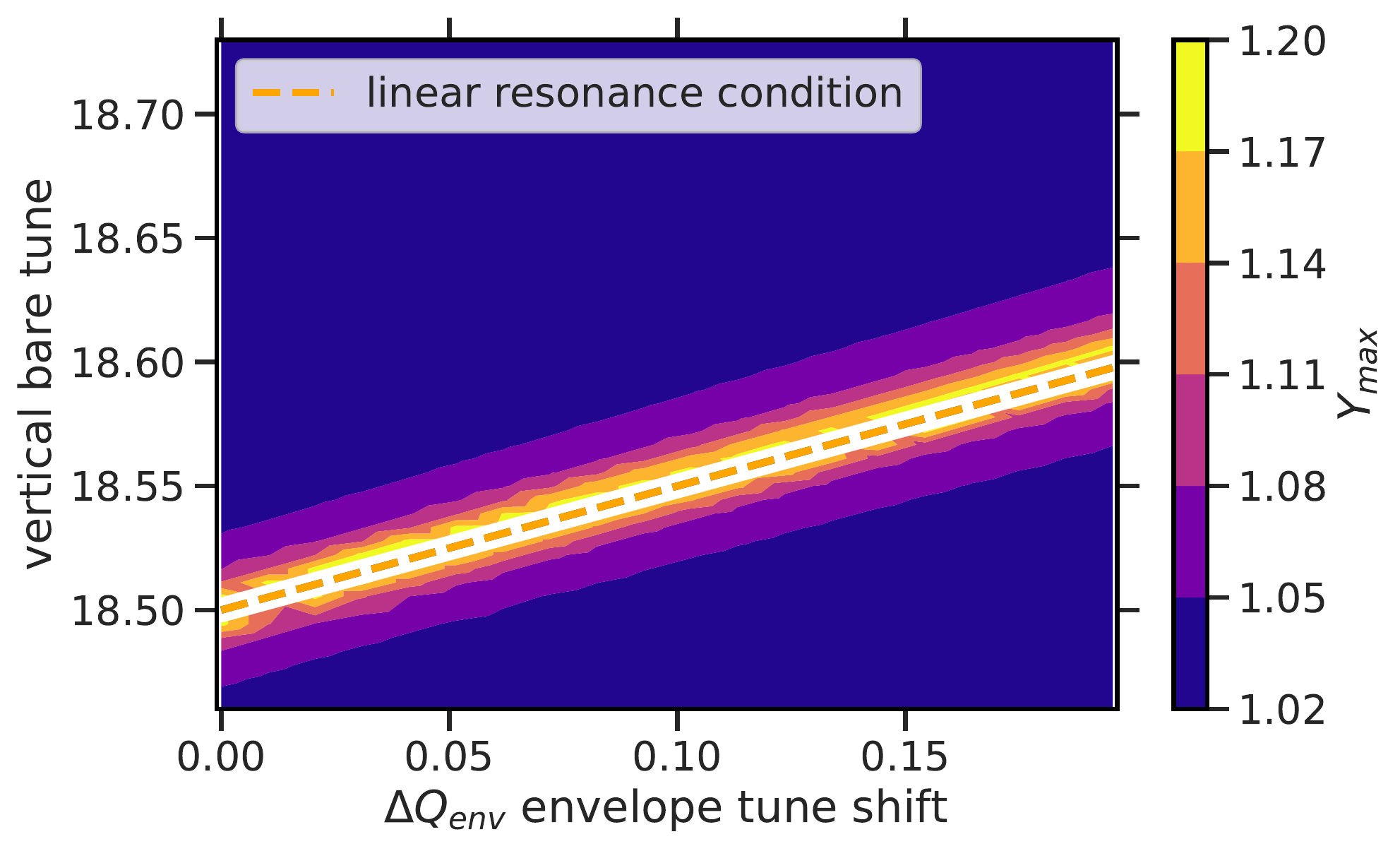}
\caption{\label{fig:kv_coasting}Solution of RMS envelope equations for the strong gradient error scenario.}
\end{figure}

RMS envelope equations~\cite{sacherer1968transverse} determine the location of the quadrupolar resonance depending on space charge for coasting beams destributed with transverse ellipsoidal symmetry. Fig.~\ref{fig:kv_coasting} illustrates the numerical solution of these equations for KV beams in terms of matched envelopes $Y_{\mathrm{max}} = \mathrm{Max}[\sigma_y / \sqrt{\epsilon_y \beta_y}]$, $\mathrm{Max}[\cdot]$ is taken over $s$.

The dark blue area depicts the resonance free areas, whereas the area with the strong deviation of the vertical envelope is shown with yellow. Note that the blue response curve in Fig.~\ref{fig:gradient_errors} is the projection of the color plot in Fig.~\ref{fig:kv_coasting} at zero space charge. The lines of the same color above and below the linear resonance condition (orange dashed line) are parallel. This indicates that the stop band width of the half-integer resonance for KV coasting beams is independent of space charge.

\subsection{\label{appendix_compensation}Correction schemes}

The goal of the optimization is to find the best set of correctors $\vec{k}^*$ minimizing the objective function $f(\vec{k}=n(\vec{\theta}))$. The vector $\vec{\theta}$ is dimensionless, and the function
\begin{equation}
n(\vec{\theta}) = \vec{k}_0 + \frac{\sqrt{Q_x^2+Q_y^2}}{R^2} \cdot N \cdot \vec{\theta}
\label{eq:normalize}
\end{equation}
normalizes $\vec{\theta}$. Here $\vec{k}_0$ corresponds to the initial settings of quadrupoles (design $k(s)$ for the main families, zero for the corrector magnets), $N$ is a diagonal matrix $N = \{... ,1/N_{k}, ...\}$, $N_{k}$ is the number of quadrupoles in the k-th family. While optimizing the SIS100 lattice different objective functions can be used. For example, it can be beta-beating for a given working point $(Q_x,\, Q_y)$. In general, there is only one condition in the choice of the objective function. It shall indicate how the quadrupolar resonance modifies for different settings of quadrupole magnets and correctors. Another consequence of changing $k(s)$ is the change of bare tunes. The constraint function $g_c(\vec{k})$ defined as
\begin{equation}
g(\vec{k}) = -\sqrt{\Delta Q_x^2(\vec{k})+ \Delta Q_y^2(\vec{k})} \geqslant 0
    \label{eq:constr_cobyla}
\end{equation}
is included into the optimization to prevent the shift of bare tunes. The process described above can be mathematically generalized using the expression
\begin{equation}
    \mathrm{Minimize}[f(\vec{k}=n(\vec{\theta})),\,\vec{\theta},\,\mathrm{constraints} = g(\vec{k})] \, .
    \label{eq:minimize}
\end{equation}

The numerical implementation is performed in Python with COBYLA \emph{Constrained optimization by linear approximation}~\cite{Powell1994}. First, the global parameters such as the integral focusing strengths of main quadrupole families and of perturbing isolated magnets (\emph{global correction scheme}) are used as optimization knobs. Second, the pair of quadrupole corrector magnets around the isolated perturbation (orange triangles in Fig~\ref{fig:survey}) is added into the optimization to suppress the gradient error locally (\emph{local correction scheme}). 

\begin{figure}[t]
\includegraphics[width=\linewidth]{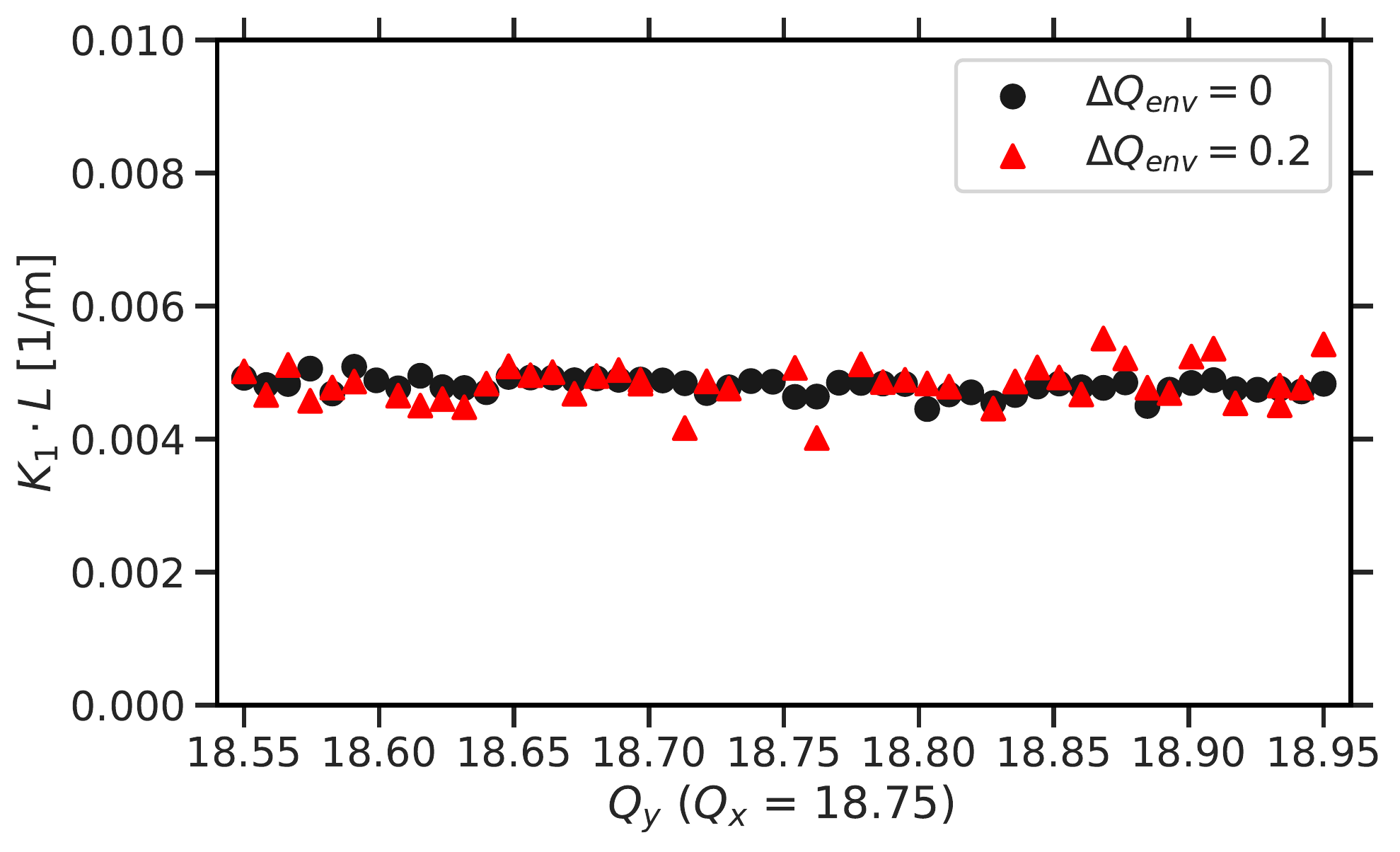}
\caption{\label{fig:corrector_values}Example of the optimal settings of the corrector magnet vs. the vertical bare tune for different values of space charge.}
\end{figure}

\begin{figure}[t]
\includegraphics[width=\linewidth]{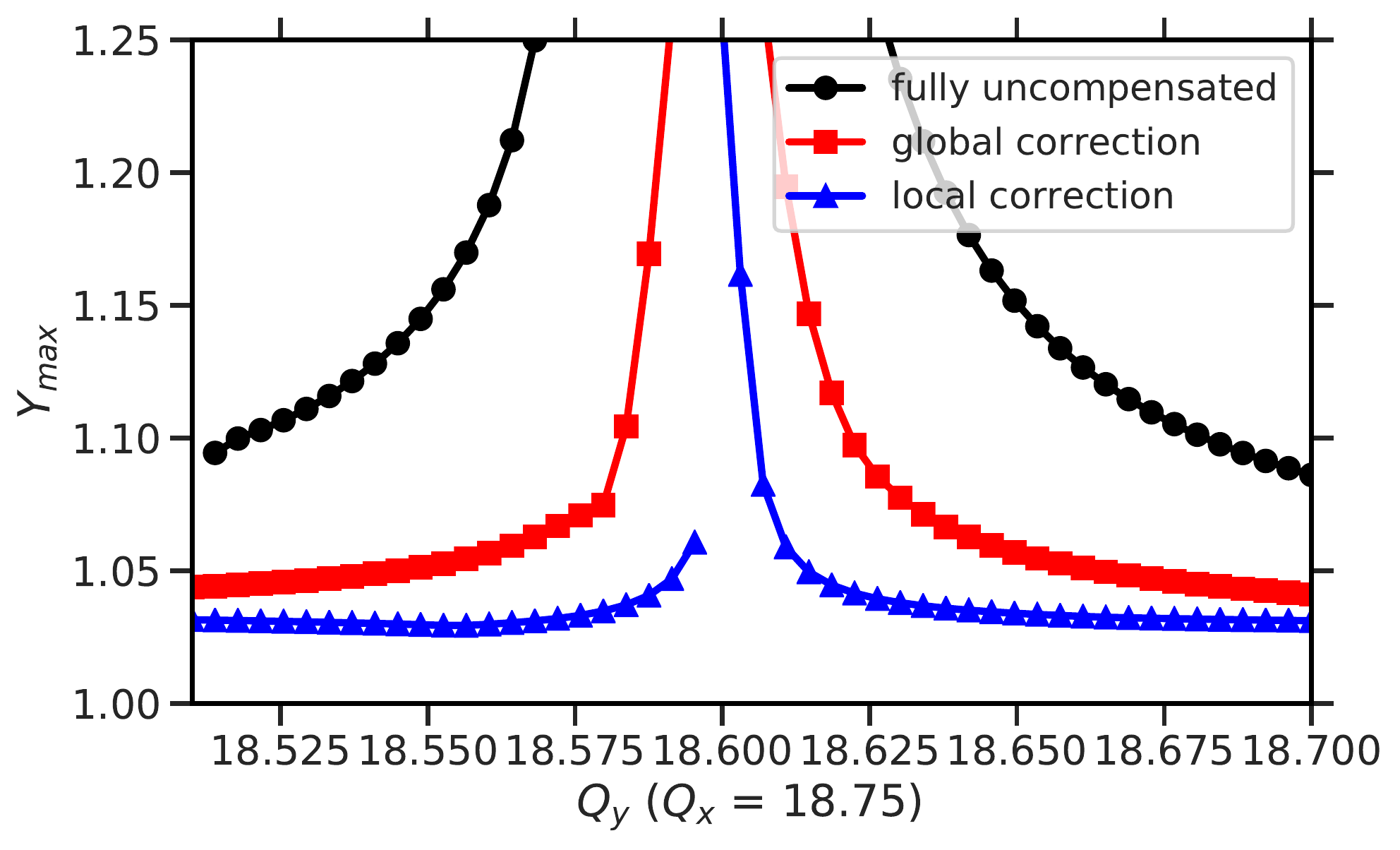}
\caption{\label{fig:after_compensation}Response of the vertical envelope to the bare tune after compensating the SIS100 lattice.}
\end{figure}

As it is shown in the Fig.~\ref{fig:edges} stronger gradient errors leading to the widening of the half-integer stop band for bunched beams. The goal of the optimization is to enlarge area of bare tunes which is free from the quadrupolar resonance. Since the horizontal stop band is also limiting the tune diagram, it should be minimized as well. Also, gradient errors and space charge modify the vertical and horizontal beam envelopes. Therefore, to keep both horizontal and vertical stop bands as small as possible the geometric sum of the horizontal and vertical beating of RMS envelopes for given bare tunes and space charge $\Delta  Q_{\mathrm{env}}$ is used as the objective function. RMS envelopes are numerically computed using full 2D envelope equations~\cite{sacherer1968transverse} with the exact lattice focusing $k(s)$ which includes all gradient errors and settings of correctors. Fig.~\ref{fig:corrector_values} shows the comparison between optimal values with and without space charge for the second adjacent corrector magnet as an example. Black dots correspond to optimal values obtained at zero space charge ($\Delta  Q_{\mathrm{env}} = 0$), whereas the red triangles are achieved at the SIS100 nominal beam parameters ($\Delta  Q_{\mathrm{env}} = 0.2$). Besides some small fluctuations, the both curves repeat each other. All magnets used in the optimization show similar behavior. This means, that optimal values obtained at zero space charge minimize the stop band width for finite space charge. Fig.~\ref{fig:after_compensation} shows the results of the half-integer resonance compensation in terms of matched beam envelopes, where the optimal parameters are found for $\Delta Q_{\mathrm{env}} = 0$.

\begin{figure}[b]
\includegraphics[width=\linewidth]{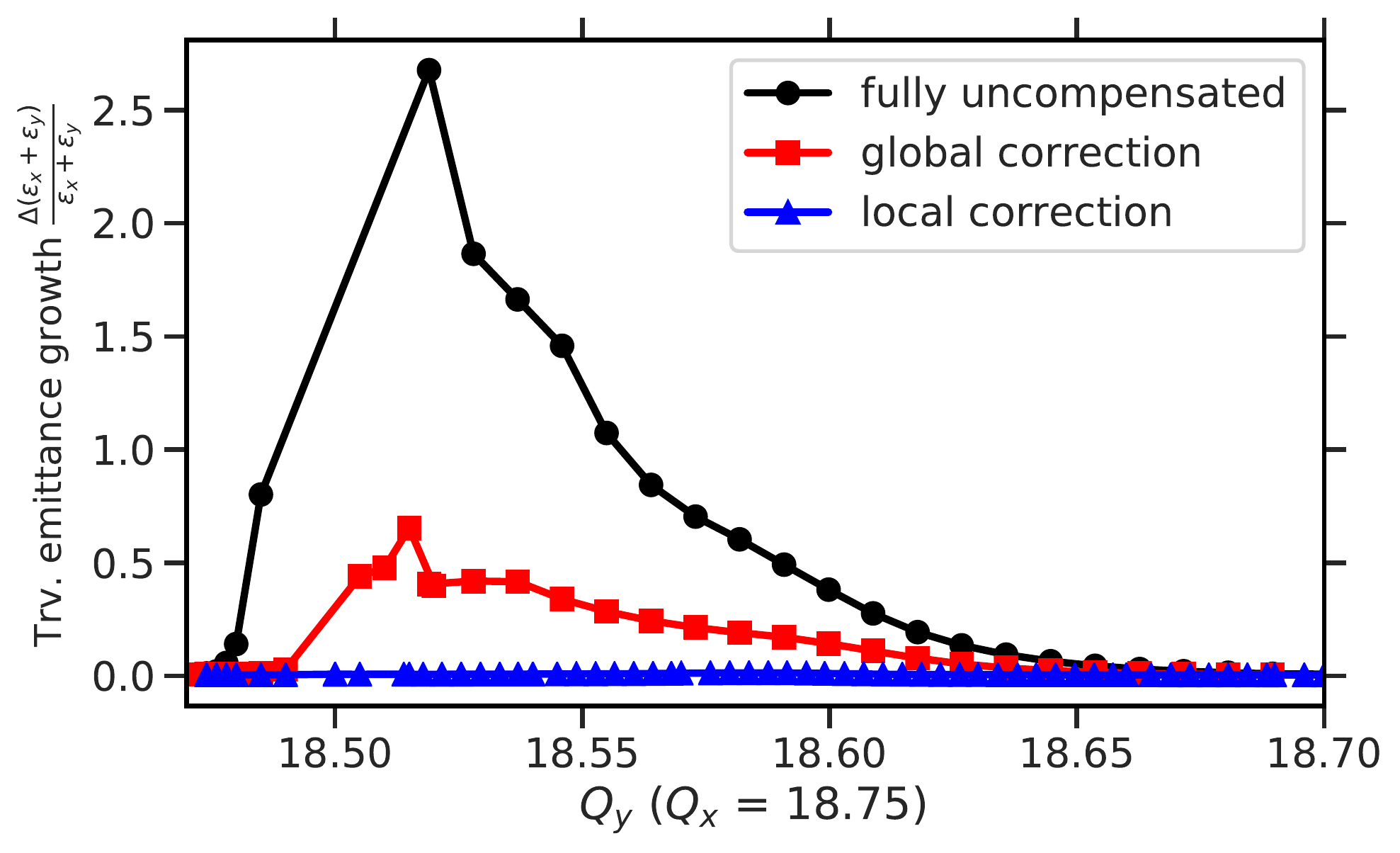}
\caption{\label{fig:full_3d_200} Transverse emittance growth of a Gaussian-like distributed bunched beam vs. the vertical bare tune after optimizing the SIS100 lattice.}
\end{figure}

Black dots here indicate the case of the initial gradient error scenario, red squares and blue triangles represent the response curves after the compensation. All curves sharply increase around the linear resonance condition at $Q_y = 18.6$, and stop band gets more narrow when more correctors are included into the optimization. Next, the results are verified using simulations for bunched beams at nominal parameters. Fig.~\ref{fig:full_3d_200} illustrates the transverse emittance growth after 200 turns.

Similarly to Fig.~\ref{fig:after_compensation} we show with black dots the initial perturbarion, red squares indicate the results of the \emph{global correction scheme}, and the blue triangles represent the \emph{local correction scheme}. The blue curve is a single line of the color plot in Fig~\ref{fig:before_after} at $Q_x = 18.75$. The optimization provided for KV beams is also valid for realistic Gaussian-like distributed bunched beams.

\subsection{\label{upper_long}Long-term dynamics and the upper edge}

\begin{figure}[b]
\includegraphics[width=\linewidth]{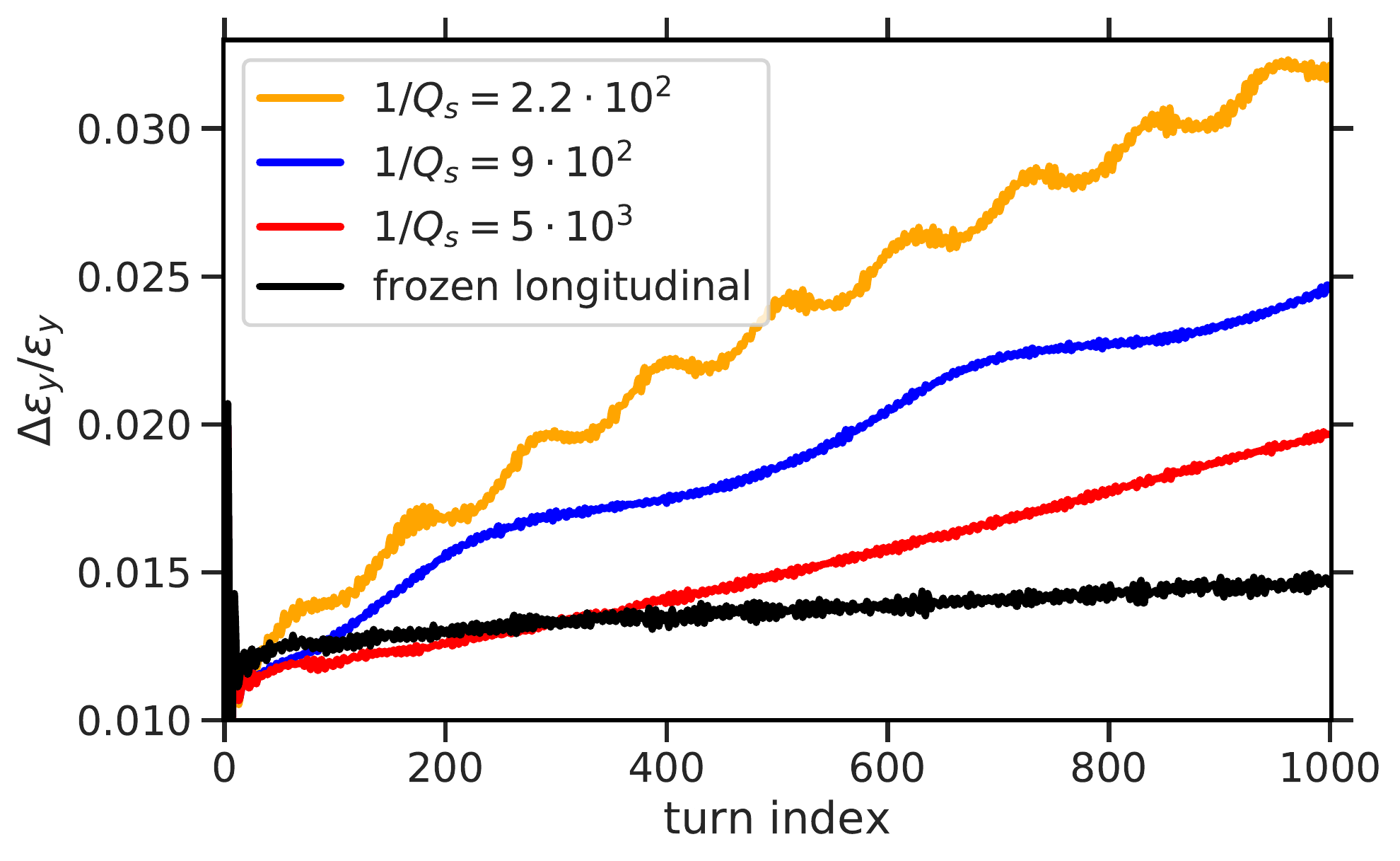}
\caption{Vertical emittance growth from simulations for different values of the synchrotron tune. The working point on $Q_y = 18.7$ corresponds to the upper edge, strong gradient error scenario.\label{fig:emittance_qs_edge}}
\end{figure}

It is important to investigate what happens with the residual emittance growth at the upper edge. The computer experiment setup developed in Sec~\ref{sec:longterm} is used. Fig.~\ref{fig:emittance_qs_edge} shows how the half-integer resonance develops in different longitudinal bunch areas at the bare tune $Q_y = 18.7$ for the strong gradient error scenario.

\begin{figure}[t]
\includegraphics[width=\linewidth]{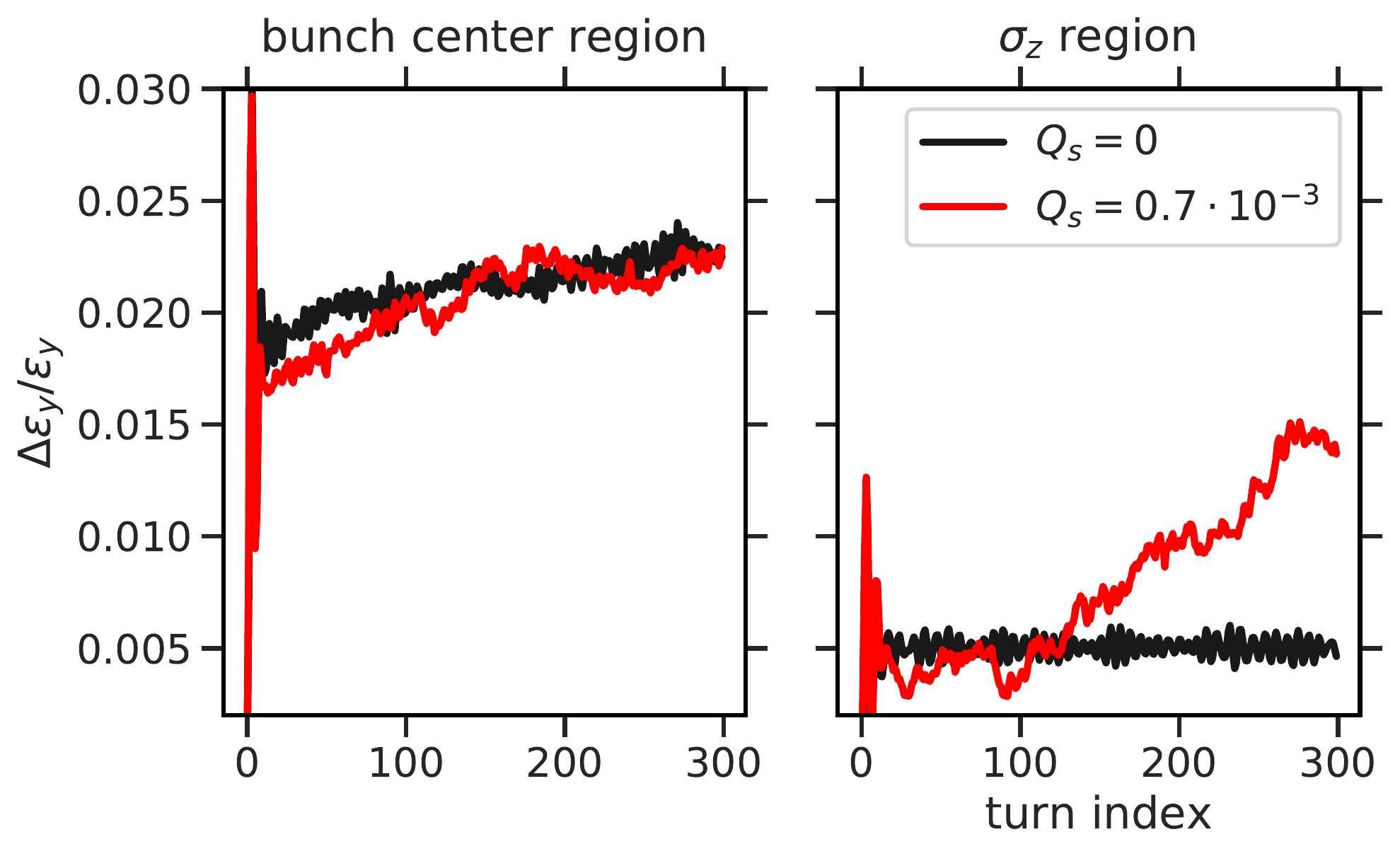}
\caption{Vertical emittance growth corresponding to separate longitudinal areas in the bunched beam for varying values of synchrontron tune\label{fig:qs_slices_edge}. The working point corresponds to the upper edge at $Q_y = 18.7$ for the strong gradient error scenario.}
\end{figure}

Black curves on both panels represent the case of frozen longitudinal motion (longitudinal coordinates $(z,\, dE)$ of all particles of the distribution remain fixed throughout the simulation), whereas reds curves show the emittance growth in the bunch center and the $\sigma_z$ region for $Q_s = 0.7 \cdot 10^{-3}$. After the initial mismatch there is some vertical emittance growth only in the bunch center region on the left. On contrary, $\sigma_z$ apart from the bunch center (on the right), the vertical emittance remains steady. Though the red curve on the left side almost repeats the black one, on the right side they are different. In the case of finite longitudinal motion, the emittance in the $\sigma_z$ region fluctuates around its initial value and starts linearly growing after 100 turns. The bunched beam keeps interacting with the half-integer resonance for any $Q_s > 0$ even at the negligibly small driving term around the upper edge. Particles gain amplitudes in the bunch center and leak towards the tails. Particles from other bunch areas start crossing the bunch center region where they can increase their amplitudes. 

Sec.~\ref{sec:longterm} shows how the emittance growth continues in the long term. The results are performed on the working point which is close to the linear resonance condition. Fig.~\ref{fig:emittance_qs_edge} illustrates what happens for different values of the synchrotron tune at the upper edge.

\begin{figure}[t]
\includegraphics[width=\linewidth]{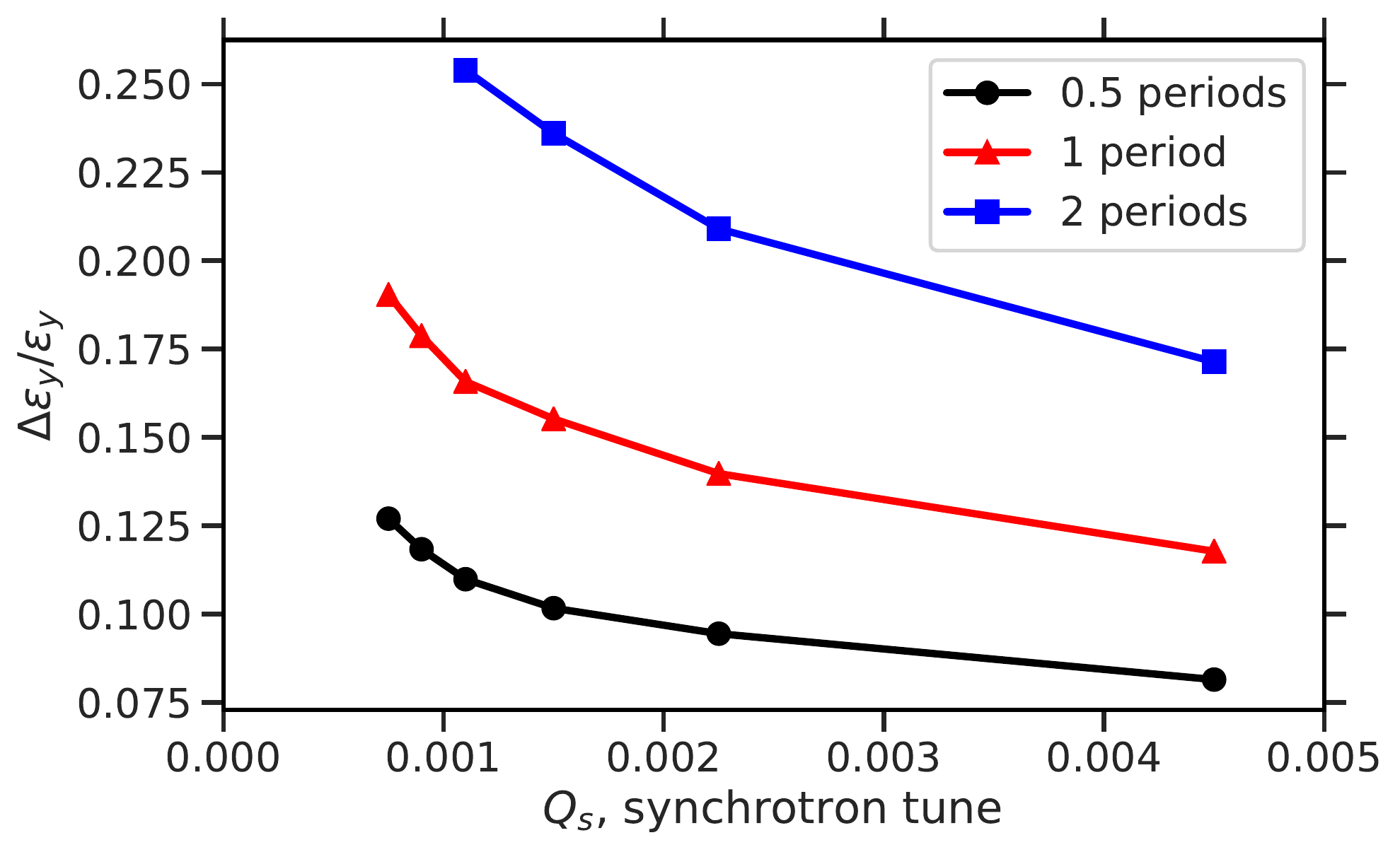}
\caption{Vertical emittance growth of a bunched beam after a few synchrotron periods. The strong gradient error is used, $Q_y = 18.62$\label{fig:qs_periods}.}
\end{figure}

The black line represents the frozen longitudinal motion. Besides the initial fluctuations and some initial growth it remains constant. With red, blue, and orange colors we show the emittance growth for the increasing value of $Q_s$. The curves oscillate on the frequency around $2Q_s$ and increase with the number of turns. Next, Fig.~\ref{fig:qs_periods} depicts how the emittance growth per a synchrotron period depends on the synchotron tune.

The results for the different number of synchrotron periods are show with different colors. All curves decrease with the increasing synchrotron tune. This can be easily explained, that for the smaller $Q_s$ the period lasts longer. Therefore, the beam accumulates more of emittance growth. The key message is that the total emittance growth per a synchrotron period is not the same for different values of $Q_s$.

\bibliographystyle{model1a-num-names}

\bibliography{bibliography}

\end{document}